\pdfoutput=1
\documentclass[prb,aps,twocolumn,superscriptaddress,showpacs,amsmath,amssymb]{revtex4}
\usepackage{graphicx}

\begin{document}

\title{Investigation of anomalous thermodynamic and transport \\ properties
   of Sr$_{1-x}$Ca$_x$RuO$_3$ ($x \geq 0.8$) }

\newcommand{\IFJ}{H. Niewodnicza\'{n}ski Institute of Nuclear Physics, Polish
 Academy of Science, Radzikowskiego 152, 31-342 Krak\'{o}w, Poland}

\newcommand{\IFUJ}{M. Smoluchowski Institute of Physics, Jagiellonian University,
  \L ojasiewicza 11, 30-348 Krak\'{o}w, Poland}

\author{A.~Zarzycki} \affiliation{\IFJ} 

\author{M.~Rams} \affiliation{\IFUJ}

\author{E. A.~G\"{o}rlich} \affiliation{\IFUJ}

\author{K.~Tomala} \affiliation{\IFUJ}

\date{May 8, 2018}



\begin{abstract}

Thermodynamic and transport properties of Sr$_{1-x}$Ca$_x$RuO$_3$ with a calcium concentration 
$x \geq 0.8$ were investigated at low temperatures and in external magnetic field. The investgations 
revealed a Landau Fermi liquid (LFL) behaviour characterized by specific heat
$C/T = const$ and resistivity $\rho\sim T^2$ and an anomalous non-Fermi liquid (NFL) behaviour with $\rho \sim T^{5/3}$ 
and a quasi-logarithmic increase of $C/T$ with decreasing temperature. The $T$--$x$ and $T$--$B$ 
phase diagrams which separate both regions were prepared for the investigated materials.  Then, 
the experimental behaviour of $C/T$ and $\rho(T)$ were compared with predictions of the self-consistent 
renormalization (SCR) theory of spin fluctuations.  Within this approach, $C/T(T)$ was well 
described up to 20 K. The SCR parameters $y_0$, $y_1$ and $T_0$ inferred from specific heat analysis allowed 
to describe properly $\rho(T)$ up to approximately 5 K.  In addition, the resistivity data were analysed within 
the `hidden Fermi liquid' theory of Anderson, obtaining very good descritption of the experimental 
behaviour up to about 25 K. An anomalous increase of the $C/T$ ratio already 
in the LFL region below 2 K in very weak magnetic fields (0.2--0.3 T) was identified as caused 
by the Schottky anomaly induced by ferromagnetic clusters embodied into the essentially 
paramagnetic materials.

\end{abstract}

\maketitle
 
\section{Introduction}
Almost to the end of the 20-th century the physical properties of correlated metals
at low temperatures were successfully described within the phenomenological Landau's
theory of the Fermi liquid (LFL).  The Fermi liquid theory assumes existence of 
the one-to-one correspondence between single particle excitations of a free electron 
gas and those of interacting electrons.  These excitations which carry the same charge, 
momentum and spin as free electrons are called quasiparticles.  The correspondence 
between the excitations spectra leads to similar behavior of thermodynamic parameters such 
as the electronic specific heat $C/T = \gamma = const$ and the magnetic susceptibility 
$\chi = const$.  The temperature dependence of the resistivity is given by 
the formula $\rho = \rho_0 + AT^2$, where $\rho_0$ is a residual resistivity.
\cite{Lohneysen07} 
Since $\gamma$ and $\chi$ are proportional to the density of  
states at the Fermi energy, both of them are proportional to an effective mass $m^*$
of the quasiparticles.  The constant $A$ in the 
resistivity formula is proportional to $(m^*)^2$.  Therefore, a linear correlation 
between $A$ and $\gamma^2$, found for a considerable group of different materials, proves legitimacy 
of this approach.\cite{Kadowaki86}  This correlation has been recently improved for heavy 
fermion cerium and ytterbium compounds, taking into account a degeneracy of the electronic 
4f shell.\cite{Tsuji05}

Nevertheless, during the last years numerous systems which show distinct deviations from the Landau's 
Fermi liquid phenomenology have been discovered.  These systems are known as non-Fermi liquid 
(NFL) materials or as the materials that show the non-Fermi liquid behaviour.  Their anomalous properties  
are often characterized by a continuous increase of the electronic specific heat coefficient
$\gamma$ with decreasing temperature, as well as by distinct 
deviation from the $T^2$ temperature dependence of the electrical resistivity.\cite{Stewart01}  
It appears that these anomalous properties are very often observed in the systems which 
are close to quantum phase transitions (QPT), i.e.\ the phase transitions that take place 
at zero temperature. In contradistinction to classical phase transitions where thermal fluctuations are 
responsible for the phase transformation, the quantum phase transitions are driven by quantum fluctuations 
of an order parameter.  Pressure, magnetic field or concentration of components in 
heterogeneous materials are usually used to tune the system to the region of QPT.  The anomalous 
properties usually cover a large part of the phase diagram making possible their experimental 
detection and characterization at higher temperatures.\cite{MVojta03}  As in the case 
of the classical phase transitions one can distinguish a first order (discontinuous) and a second 
order (continuous) transitions.

First of all, the investigations of the NFL properties concentrated mostly on the heavy fermion 
materials in the vicinity of the QPT between an antiferromagnetic (AF) and a paramagnetic (PM) states. 
 As the best studied examples one should mention CaCu$_{6-x}$Au$_x$ system 
with $x \simeq 0.1$ where the pressure, the concentration of components or the external magnetic 
field were applied as tuning parameters \cite{Lohneysen94} and YbRh$_2$Si$_2$ which shows the AF order
with the N\'{e}el temperature of $T_N \simeq 70$~mK but can be tuned to the quantum critical point
by a relatively weak external magnetic field.\cite{Trovarelli00} Both systems at low 
temperatures show the anomalous behaviour of the specific heat $C/T \sim -\log T$ and the resistivity 
$\varrho \sim T$ over more than a decade of temperature.

The important group of materials in which the magnetically ordered state can be tuned to the paramagnetic 
state by variations of the pressure or the composition is formed by weak itinerant ferromagnets.  Among 
the numerous investigated materials one can mention the probably most representative clean systems: 
MnSi,\cite{Pfleiderer97} ZrZn$_2$,\cite{Uhlarz04} and NiAl$_3$\cite{Niklowitz05} as well as the systems 
with a quenched disorder: Ni$_{1-x}$Pd$_x$,\cite{Nicklas99} Zr$_{1-x}$Nb$_x$Zn$_2$,\cite
{Sokolov06} and Ni$_3$Al$_{1-x}$Ga$_x$.\cite{Yang11}.  The Sr$_{1-x}$Ca$_x$RuO$_3$ system, which is 
the subject of studies presented in this paper, can be included in the last group.  Recently, the results of  
investigations of the metallic systems being on the border
of itinerant magnetism and metallic paramagnetism 
were gathered and discussed in the extensive review paper.\cite{Brando16}      

The magnetic properties of the Sr$_{1-x}$Ca$_x$RuO$_3$ compounds were already seriously 
investigated and a significant number of publications have appeared up to now.
\cite{Kanbayasi78,Cao97,Kiyama97,Yoshimura98,Kiyama99,Fuchs15} 
The system shows a complete solid solubility for the entire range of concentrations.  
The increase of the calcium concentration is connected with increase of the orthorhombic 
GdFeO$_3$ type distortion of the cubic perovskite lattice leading to increased buckling of the 
Ru--O--Ru bond which is probably responsible for the magnetic behaviour.\cite{Fang02}   
The pure strontium  and calcium  compounds show distinctly different magnetic behaviour.  SrRuO$_3$ is 
a metallic ferromagnet with the Curie temperature $T_C\simeq 160$~K, a paramagnetic Curie temperature 
$\Theta_p\simeq 160$~K and a spontaneous magnetic moment $\mu_s\simeq0.84\mu_B$ per Ru atom, 
the last one being rather far from the expected magnetic moment of the Ru$^{4+}$ ion with a 4d$^4$ 
configuration in a low spin state with $S = 1$.\cite{Kiyama99}  
Substitution of Sr$^{2+}$ ions by smaller Ca$^{2+}$ leads to the gradual decrease of the Curie 
temperatures and simultaneously to decrease of the ruthenium magnetic moments.  For a long time there 
was common consent that the ferromagnetic order disappears at some 
critical concentration near $x_{cr}\simeq0.7$ that was obtained by an extrapolation of the 
concentration dependence of $T_C(x)$ for the compounds on the strontium rich side.  
For Sr$_{0.4}$Ca$_{0.6}$RuO$_3$, classified still 
as a homogenous ferromagnet (for an explanation see below), $T_C\simeq 25$~K and 
$\mu_s\simeq0.16\mu_B$ per Ru atom.  The temperature dependence of the magnetic 
susceptibilities at high temperatures (approximately above 60--80 K),  for all Sr$_{1-x}$Ca$_x$RuO$_3$ 
compounds can be well described by the Curie-Weiss formula.  Nevertheless, drastic decrease of 
the paramagnetic Curie temperatures $\Theta_p$ is observed which for materials with $x\gtrsim0.6$ become 
negative.  However, the paramagnetic magnetic moments  
are relatively constant and show only a minor increase from $\mu_{eff}\simeq2.7\mu_B$ in SrRuO$_3$ 
to above $\mu_{eff}\simeq3.0\mu_B$ for calcium rich materials.  In spite of the very large 
negative paramagnetic Curie temperature $\Theta_p\simeq -150$~K which could suggest the 
antiferromagnetic order and $\mu_{eff}\simeq 3.4\mu_B$ at high temperatures, CaRuO$_3$ does not 
order magnetically.\cite{Yoshimura98,Kiyama99,Mukuda99,Koriyama04,Rams09}
It was classified as a strongly exchange enhanced paramagnet 
(Stoner enhancement factor $\alpha\simeq 0.97$) on the border of ferromagnetism.
\cite{Mukuda99}  The magnetic investigations performed in our Laboratory and shown in the Supplemental 
materials allow to conclude that the Sr$_{1-x}$Ca$_x$RuO$_3$ compounds with the calcium concentrations 
$x \geqslant 0.8$ are essentially paramagnetic.  

This short review shows that the magnetic properties of Sr$_{1-x}$Ca$_x$RuO$_3$ cannot be 
described as ordering of the magnetic moments localized at ruthenium atoms.  They show the distinct 
features of the itinerant magnetism and the zero temperature  transition around $x_{cr}\simeq 0.7$ 
occurs between the itinerant ferromagnet and the paramagnetic metal.

Important information about the magnetic behaviour of Sr$_{1-x}$Ca$_x$RuO$_3$ and especially about 
the concentration tuned quantum phase transition  were inferred 
from the precise magnetisation measurements using a magneto-optical Kerr rotation method
for a thin epitaxial film with practically continuous variation of the calcium 
concentration. \cite{Demko12}  It has appeared that the strong crystallographic distortion 
caused by different radii of the Sr and Ca atoms leads to significant extension of the ferromagnetic 
phase over a wide range of compositions $x$ which can be understood as caused by formation of an inhomogenous 
ferromagnetic state built up of frozen magnetic clusters embodied into the paramagnetic metal.  
There is no any abrupt change of the Curie temperature on the $T$--$x$ phase diagram, 
which means that the composition tuned quantum transition between the ferromagnetic and 
paramagnetic phases in Sr$_{1-x}$Ca$_x$RuO$_3$ is smeared and practically destroyed by the strong 
crystallographic disorder.  It means that instead of a single critical concentration 
$x_{cr}\simeq 0.7$ there exists rather a critical range of concentrations within which the transition  
takes place.  
In the following parts of the text we sometimes use a notion `critical concentration of the phase 
transition'  (usually written in the quotation mark), nevertheless, one should always understand it as a 
critical range of concentrations.

In order to finish this short review of the magnetic properties of Sr$_{1-x}$Ca$_x$RuO$_3$ perovskites 
one should also consider the  results of the investigations using different microscopic methods like 
muon spin rotation  ($\mu$SR),\cite{Uemura07,Gat11} $^{16}$O and $^{99}$Ru nuclear magnetic 
resonances (NMR)\cite{Yoshimura98,Mukuda99} and neutron scattering (NS).\cite{Gunasekera15}  

A very important piece of information provided by the $\mu$SR experiments concerns the existence 
of a phase separation in Sr$_{1-x}$Ca$_x$RuO$_3$ for compositions with the calcium concentration $x \geq 0.65$, just around 
the recently discussed `critical concentration' at $x_{cr}\simeq 0.7$. The compounds in 
this range of concentrations are magnetically inhomogenous and 
contain both ferromagnetic and paramagnetic fractions.  This result is in agreement with the suggestion of the 
inhomogenous magnetic state inferred from the thin film 
magnetisation measurements.\cite{Demko12}  Both NMR and NS are sensitive to 
the dynamics of the magnetic moments.  The NMR investigations allowed to discover very strong ferromagnetic 
spin fluctuations for Sr$_{1-x}$Ca$_x$RuO$_3$ compounds in the whole range of calcium concentrations including pure 
CaRuO$_3$, in spite of the large negative paramagnetic Curie temperature in this compound.\cite{Yoshimura98}  
The spin fluctuations in CaRuO$_3$ were also discovered by the inelastic neutron scattering.
\cite{Gunasekera15}  In addition, detailed analysis of the neutron scattering profile suggests 
`the formation of small ferromagnetic domains, behaving as dynamic paramagnetic clusters of Ru$^{4+}$
spins'.

An important difference between strontium and calcium compounds concerns not only the magnetic behaviour but also 
their thermodynamic and transport 
properties.  SrRuO$_3$ demonstrates the LFL behaviour with $\rho \sim T^2$ at low temperatures 
\cite{Capogna02,Khalifah04} and $C/T = const$ with the enhanced electronic specific heat coefficient 
$\gamma \simeq 30$~mJ/molK$^2$.\cite{Allen96,Kiyama98,Cao97,Khalifah04,Cao08}  The LFL 
 properties are preserved 
after substitution of Sr$^{2+}$ ions by Ca$^{2+}$ at least to the calcium concentration $x \simeq 
0.6$, that means for samples which are ferromagnetically ordered.\cite{Khalifah04}  However, the paramagnetic
samples with the calcium concentration above approximately $x = 0.8$ demonstrate the NFL 
properties characterized by  a temperature dependence of the resistivity different from $T^2$ and often anomalous
behaviour of $C/T$ at low temperatures.\cite{Kiyama98,Khalifah04}  
The most frequently investigated pure CaRuO$_3$   
shows NFL features both in the resistivity and the specific heat.  The results of the 
resistivity measurements performed on single crystals, thin films and polycrystalline samples above 
2 K point to the temperature dependence $\rho \sim T^{3/2}$.\cite{Klein99,Capogna02,Khalifah04}
Only recently, the LFL state was found in CaRuO$_3$ below 1.5 K with recovery of the anomalous behaviour  
at higher temperatures.\cite{Schneider14}  The results of the specific heat measurements 
of CaRuO$_3$ demonstrate at low temperatures a wide variety of behaviour.  Some investigations show only 
the LFL properties and the specific heat is well described by the linear electronic contribution and 
a lattice contribution given by the 
Debye model.\cite{Cao97,Shepard97,Kikugawa09}  The other investigations show the anomalous NFL 
upturn of $C/T$ at low temperatures which was analysed within the spin-fluctuation theory of Moriya (see below), 
\cite{Kiyama98} the upturn with some mysterious and magnetic field 
dependent structure around 2 K,\cite{Cao08} and the upturn with the power law temperature dependence 
down to 100 mK.\cite{Baran12}  Recently, the continuous increase of $C/T$ with decreasing temperature 
was reported down to 1.8 K, nevertheless, without any theoretical analysis of $C/T$  
for materials with $x \geq 0.7$.\cite{Fuchs15}

Simultaneously with the experimental investigations theoretical methods which could allow to understand 
and describe the deviations
from the Landau Fermi liquid behaviour were developed.  Since these anomalous behaviours often take  
place close to the zero temperature phase transitions which are caused by strong quantum 
fluctuations of the order parameter it suggests that their theoretical description should be connected with 
the theory of the quantum phase transitions.  
The first theoretical analysis of the zero temperature phase transition from the itinerant ferromagnet to
the paramagnetic metal was performed within the $\phi^4$ quantum field theory with the 
Ginzburg-Landau-Wilson functional which was analysed by the renormalization group (RG) 
method.\cite{Hertz76}  It was found that the quantum phase transition for a system with a spatial dimension $d$ can 
be analysed as classical transition for the system with the effective dimension $d_{eff} = d + z$, where 
$z$ is a dynamic critical exponent.  In the next step, the theoretical investigations within this 
framework were extended to higher temperatures and allowed to make predictions 
about the temperature behaviour of different physical parameters in the critical region for 
the materials which show FM and AFM orders and of different spatial dimensionality.\cite{Millis93}  For example, 
for the transition between the 3-dimensional itinerant ferromagnet and paramagnetic metal the heat capacity should 
scale as $C/T \sim -\log T$ and the temperature dependence of the resistivity should behave as 
$\rho(T) \sim T^{5/3}$.  Independently, another theoretical approach used to describe  
the properties of the itinerant magnets close to the border of the magnetic instability was developed
.\cite{Moriya85}  This approach stresses importance of the spin fluctuations both in nearly magnetic and magnetically 
ordered FM\cite{Moriya73,Ishigaki96} or AFM\cite{Moriya95} states.  The theory, called the 
self-consistent renormalization (SCR) 
theory of spin fluctuations, predicts the phase diagrams and the temperature variation of many 
physical parameters including the magnetic susceptibility, specific heat and electrical resistivity.  
The same kind of the spin fluctuation theory was independently developed by another authors.
\cite{Lonzerich85} The results of the theoretical investigations in the critical region of the 
magnetic instabilities were gathered in the already quoted extended review article.\cite{Stewart01} 
 
Here we report the results of the zero field and magnetic field tuned specific heat 
and electrical resistivity investigations of Sr$_{1-x}$Ca$_x$RuO$_3$ compounds which are in the 
paramagnetic phase above the zero temperature phase transformation.  Since in the region around 
the discussed `critical concentration' ($x \simeq 0.7$) the compounds are magnetically inhomogenous 
we confined the studies to the materials with the compositions $x = 0.8$, 0.9, and 1.0.  They are 
still quite close to the region of concentrations where the ferromagnetic to paramagnetic 
transformation takes place at zero temperature but simultaneously they are homogenously  
paramagnetic.  The conclusion about their paramagnetic behaviour can be drawn from the results 
of the bulk magnetic measurements presented in Supplemental materials where the results of investigations 
performed for the compounds 
within the extended range of compositions from $x = 0.60$ up to $x = 1.0$ are reported.  They prove that 
in general the materials 
with the calcium concentration $x \geq 0.8$ are paramagnetic, at least down to 2 K.\cite{SupplMat}  
These do not exclude a possible existance of a small amount of ferromagnetic clusters which can strongly
influence the magnetic properties at low temperatures.
Since no any irregularities in the temperature dependences of the specific heat 
down to 0.4 K and the electrical resistivity down to approximately 0.7 K were detected it seems that 
they do not show any  magnetic order to much lower temperatures. The essentially paramagnetic behaviour of the 
$x = 0.8$ sample is also strongly supported by the results of a $^{99}$Ru M\"{o}ssbauer effect investigations 
which are shown in the Supplemental materials.

The experimental results were analysed in a few ways.  First of all, the Fermi liquid region 
was identified using $C/T$ and $\rho$(T) behaviour and the phase diagrams on the $T$--$x$ and $T$--$B$ 
planes were prepared, which demonstrate the regions of FL and NFL properties.  Then, the results of $C/T$ 
and $\rho(T)$ were compared with the predictions of the Moriya's SCR theory of spin fluctuations 
\cite{Moriya85,Ishigaki96} and with the `hidden Fermi liquid' theory of Anderson.\cite{Anderson08}
At the end, we provide a summary of the experimental results and present the conclusions inferred 
from their analysis.

\section{Experimental details}
The polycrystalline Sr$_{1-x}$Ca$_x$RuO$_3$ compounds with a few calcium concentrations between 
$x = 0.60$ and 1.0 were prepared by the solid state reaction from weighted in a proper 
molar ratios RuO$_2$, SrCO$_3$ and CaCO$_3$.  First of all, mixed and pressed powders were calcined for 
several hours at 800$^\circ$C.  Then, after grinding, they were pressed into pellets under high 
pressure and sintered  at high temperatures for approximately 15 hours.  The compounds with the 
calcium concentrations between  $x = 0.60$ and 0.80 were sintered at 1300$^\circ$C and the grinding procedure 
was repeated several times (at least 3 times).  The materials with $x = 0.9$ and 1.0 were sintered 
in sequence at 1000$^\circ$C, 1100$^\circ$C and 3 times at 1200$^\circ$C.  All heat treatments were performed in the atmosphere of flowing Ar + 1$\%$O$_2$.  After sintering the samples were cooled down with the furnace.

The crystal structures of the prepared materials were verified by the X-ray diffraction using the 
Cu K$_{\alpha}$ radiation and D501 Siemens powder diffractometer.  The diffraction patterns were 
analysed using the FULLPROF program within the $Pnma$ space group, which describes 
the perovskite structure with orthorhombic distortion of the GdFeO$_3$ type. It was found that 
all synthetized materials were single phase.

Due to the observation of the smeared zero temperature ferromagnetic to paramagnetic phase transition 
and the inhomogenous magnetic state in the critical region discovered 
by the $\mu$SR measurements\cite {Uemura07,Demko12} the investigations of the bulk magnetic properties 
were used to select materials which being close to the region of the phase 
transformation were still paramagnetic. The magnetization measurements for the compounds within the 
concentration range from $x = 0.6$ up to 1.0 were performed using MPMS XL5 SQUID magnetometer from Quantum Design
including low field susceptibility ($B \simeq 5$~mT), ac susceptibility, hysteresis loop at 2 K and study of the 
magnetic equation of state (Arrott plot).  These measurements allowed to conclude 
that the compounds with the calcium concentrations $x \geq 0.8$ are essentially paramagnetic, at least down to 
2 K. The results of these investigations are presented in the Supplemental materials.  
The Supplemental materials contain also the results of $^{99}$Ru M\"{o}ssbauer 
effect investigations of the compounds with compositions $x = 0.6$ and 0.8.  They confirm the paramagnetism 
of the $x = 0.8$ compounds (at least down to 4.2 K) and the homogenous ferromagnetism of the $x = 0.6$ material 
where homogenous means that all ruthenium magnetic moments take part in the ferromagnetic ordering.
Taking into account the results of all these investigations, Sr$_{1-x}$Ca$_x$RuO$_3$ compounds with 
the compositions $x = 0.8$, 0.9, and 1.0 were selected for further studies of the thermodynamic and 
transport properties.

The specific heat and the electrical resistivity measurements were performed using the PPMS equipment 
from Quantum Design with the $^3$He option and 9~T superconducting magnet.  This equipment in our case allowed to 
carry out the specific heat measurements with the required accuracy in the range of temperatures down 
to 0.4~K in the external magnetic fields up to 3~T. The Apiezon N grease was used to fix the samples in 
the microcalorimeter. Its heat capacity was measured for each sample at each magnetic field, and then subtracted. 
The smooth calibration of microcalorimeter thermometers occured to be very important to obtain reproducible and
smooth specific heat dependences.

The dc current option of the the resistivity measurements allowed to perform 
the measurements down to 0.6--0.7~K in the full range of the accessible magnetic fields. 
  
\section{Experimental results and analysis}
In this section, the results of the investigation of the specific heat and the electrical resistivity 
are presented.  In the 
first part, the zero field measurements are shown and used to prepare a phase diagram which separates on the 
$T$--$x$ plane the regions of FM, LFL and NFL properties.  The LFL region was defined by the 
Landau Fermi liquid temperature $T_{LFL}$ 
derived from the upper limit of $C/T = const$ behaviour in the specific heat and from the
upper limit of the $\rho \sim T^2$ 
temperature dependence of the resistivity.  Then, the phase diagrams on the $T$--$B$ plane for all 
the investigated compounds are prepared  
from the results of the specific heat and resistivity measurements in the external magnetic field.
Since the dc current method of the resistivity measurements used in our PPMS system has 
not enough accuracy to identify without any doubt the upper limit of $\rho \sim T^2$ dependence 
below approximately 2 K (an exception is the $\rho(T)$ dependence 
in CaRuO$_3$ at zero field) in this range of temperatures only the results of the specific heat 
measurements were used to identify the LFL properties. On the contrary, in the external magnetic 
fields above 3 T when the LFL region growth to much higher temperatures the  
$T_{LFL}$ temperature can be determined from the results of the resistivity measurements.  In the 
NFL region the specific heat follows the quasi-logarithmic behaviour and the 
electrical resistivity was analysed using the  $\rho \sim T^{5/3}$ relation.  

\begin{figure}[htb!]
\includegraphics[width=8cm]{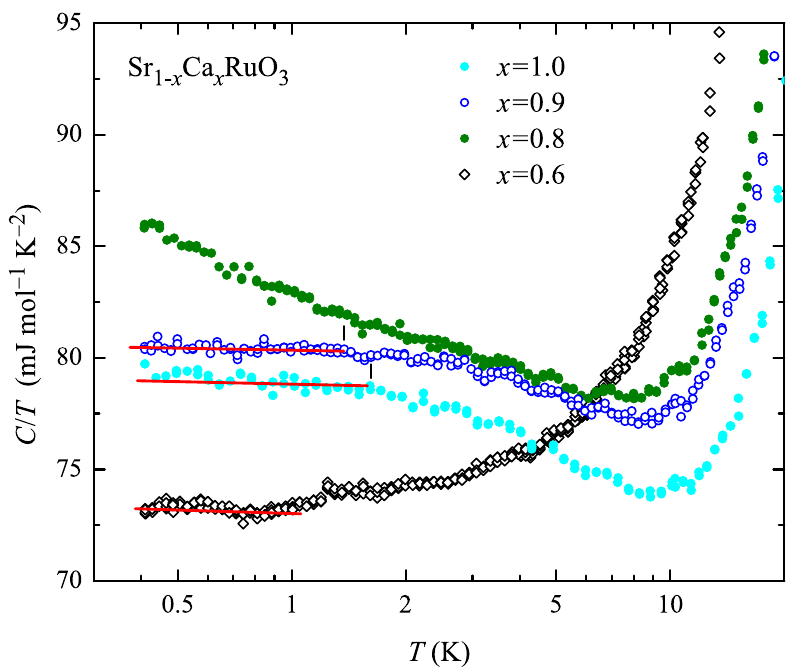}
\caption{\label{f1} 
(Color online) Temperature dependence of specific heat $C$ measured in zero magnetic field
for several compositions of Sr$_{1-x}$Ca$_{x}$RuO$_3$. The constant $C/T$, as marked by red lines, 
indicate LFL behaviour.}
\end{figure}

\begin{figure}[htb!]
\includegraphics[width=8.5cm]{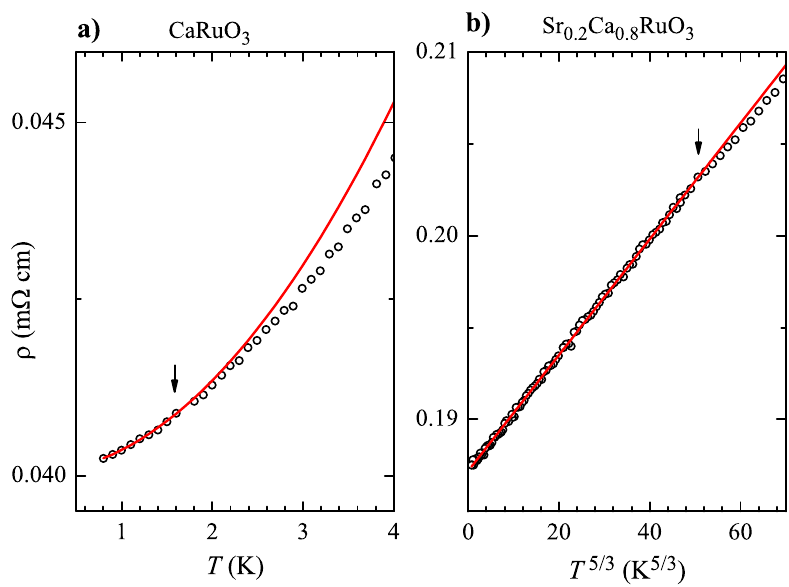}
\caption{\label{f2} (Color online) Temperature dependencies of electrical resistivity for CaRuO$_3$ (a) and 
for Sr$_{0.2}$Ca$_{0.8}$RuO$_3$ (b). The solid lines were fitted using $T^2$ and $T^{5/3}$ laws,
respectively.}
\end{figure}

\subsection{Zero field specific heat and resistivity}
The temperature behaviour of the specific heat presented as $C/T$ versus $\log T$ for the compounds with 
the calcium concentrations $x = 0.6$, 0.8, 0.9, and 1.0 is shown in Fig.~\ref{f1}.  At first, one can directly 
confirm that the ground states of Sr$_{0.1}$Ca$_{0.9}$RuO$_3$ and CaRuO$_3$ are the Landau Fermi 
liquids.  In both materials the $C/T = const$ behaviour extends above 1 K.
Identification of the LFL ground state in CaRuO$_3$ below $T_{LFL} = 1.60(15)$~K is in good agreement 
with the recently published results of the resistivity measurements $T_{LFL} = 1.5$~K\cite{Schneider14} 
and with the results of our transport investigations which give $T_{LFL} = 1.6(2)$~K (see below).   
The temperature dependence of specific heat 
for Sr$_{0.2}$Ca$_{0.8}$RuO$_3$ is distinctly different and $C/T$ increases  
to the lowest accessible temperature. The reason of this anomalous increase will be discussed 
in one of the following sections.  However, one cannot exclude a recovery of the LFL state at lower 
temperatures (see the results of the specific heat measurements in external magnetic fields shown in 
the next section), nevertheless, it could appear only below approximately 0.5 K.     
For comparison, we present also the results of the specific heat measurements for 
the $x = 0.6$ sample which is the homogenous ferromagnet and shows the LFL properties.
\cite{Uemura07,Khalifah04}

\begin{figure}[htb!]
\includegraphics[width=8cm]{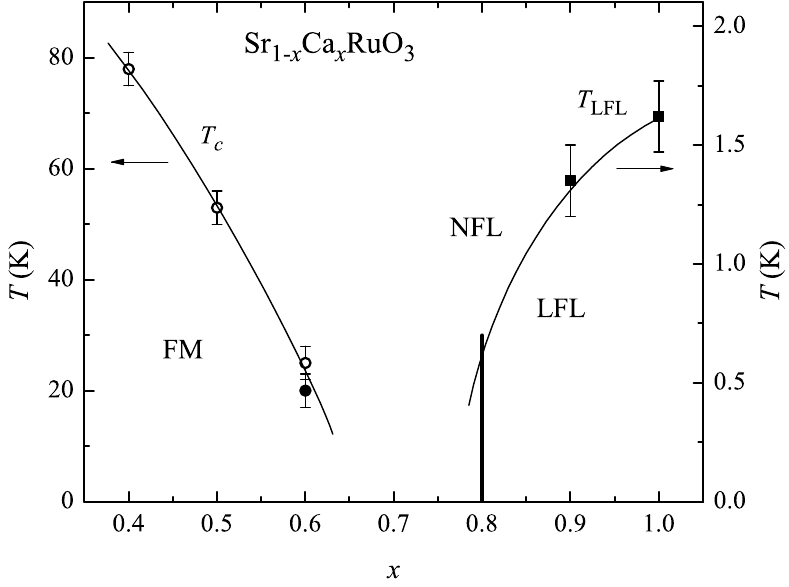}
\caption{\label{f3} 
$T$-$x$ phase diagram for Sr$_{1-x}$Ca$_{x}$RuO$_3$ with regions of ferromagnetic (FM), Landau Fermi liquid (LFL) and non-Fermi liquid (NFL) behaviour.}
\end{figure}

The conclusions inferred from the temperature variations of the specific heat are corroborated 
by the results of the electrical resistivity measurements.   In Fig.~\ref{f2} the zero field  
of $\rho(T)$ dependence for CaRuO$_3$ and Sr$_{0.2}$Ca$_{0.8}$RuO$_3$ are shown.  
For CaRuO$_3$ is well described by the $\rho(T) = \rho_0 + a_2 T^2$ formula.  This formula fitted to experimental points 
below 1.6 K is in Fig.~\ref{f2}a extended to higher temperatures 
to show deviation from the $T^2$ law above 1.6 K.  The temperature dependence of the resistivity for 
Sr$_{0.2}$Ca$_{0.8}$RuO$_3$, shown in Fig.~\ref{f2}b, can be well analysed using the NFL formula 
$\rho(T) = \rho_0 + a_{5/3} T^{5/3}$ to the lowest accessible in our experiment temperature of 
0.7~K. 

The experimental results allow to prepare the phase diagram on the $T$--$x$ plane which indicates 
FM (for simplicity only the values of $T_C$ for the compounds with the calcium concentrations 
$x = 0.4$, 0.5 and 0.6 are shown), LFL and NFL phases (Fig.~\ref{f3}).  

\begin{figure}[htb!]
\includegraphics[width=8cm]{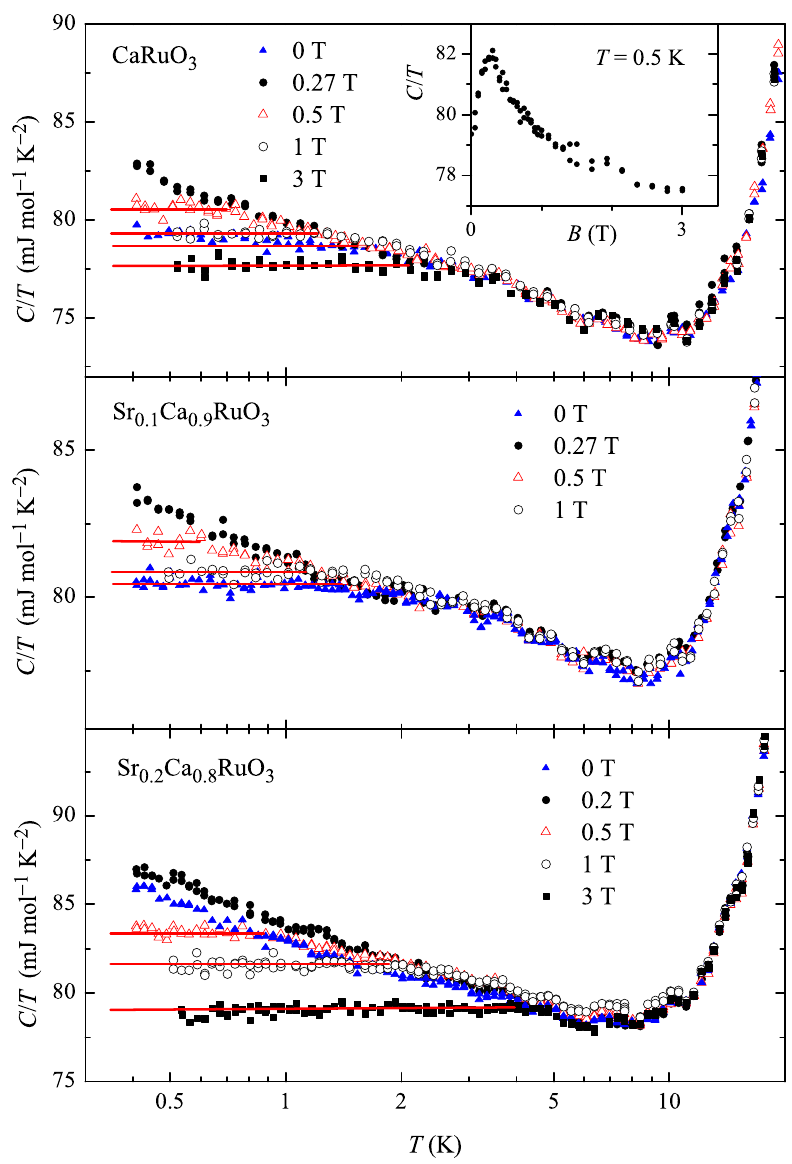}
\caption{\label{f4} 
(Color online) 
Temperature dependence of specific heat $C$ measured at different magnetic fields
for CaRuO$_3$, Sr$_{0.1}$Ca$_{0.9}$RuO$_3$, and Sr$_{0.2}$Ca$_{0.8}$RuO$_3$. The solid red lines indicate the regions of LFL behaviour. Inset: Magnetic field dependence of $C$ for CaRuO$_3$ at 0.5 K.}
\end{figure}
  
\subsection{Specific heat and resistivity in external magnetic field}
The specific heat behaviour in different external magnetic fields, presented as $C/T$ on the half-logarithmic 
temperature scale, are shown for all investigated compounds in Fig.~\ref{f4}. 
At first, one should notice as a general feature that the most distinct increase 
of $C/T$ dependence to the lowest accessible temperatures take place not in zero but in very weak 
critical magnetic fields which amount to approximately $B_0 \simeq 0.27(2)$~T in CaRuO$_3$ and 
Sr$_{0.1}$Ca$_{0.9}$RuO$_3$ and $B_0\simeq 0.20(2)$~T in Sr$_{0.2}$Ca$_{0.8}$RuO$_3$.  
The values of these critical fields could be identified from the magnetic field dependence of $C/T$ at 0.5 K 
which have the distinct maxima at $B_0$ shown for the $x = 1.0$ compound in the insert 
in Fig.~\ref{f4}.   In the magnetic 
fields which are lower (in our case it is only $B = 0$) or higher  than these critical values the LFL 
properties are well seen and one can notice a gradual increase of the regions of the Fermi liquid 
behaviour with increasing field.  In Sr$_{0.2}$Ca$_{0.8}$RuO$_3$ the temperature dependence of $C/T$ shows 
the distinct increase to the lowest temperatures in the field of 0.2~T.  These anomalous increase 
of the $C/T$ ratio in the weak magnetic fields are discussed separately.

The values of $T_{LFL}$ derived from the upper limit of the $C/T = const$ behaviour were used to prepare the 
low temperature part of the phase 
diagrams which separate regions with LFL and NFL behaviour on the $T$--$B$ plane (Fig.~\ref{f7}).  
The numerical values of $T_{LFL}$ are gathered in Table I in the Supplemental materials.

\begin{figure}[htb!]
\includegraphics[width=8.5cm]{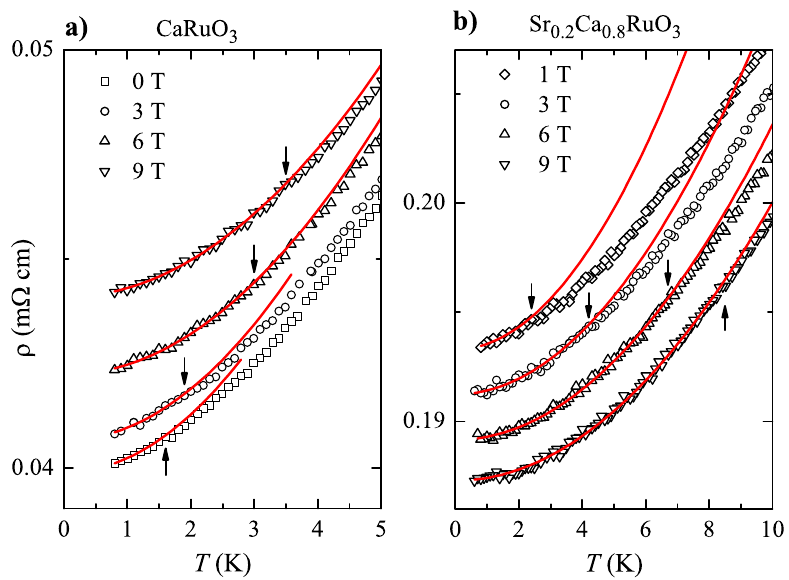}
\caption{\label{f5} 
(Color online) 
Resistivity of a) CaRuO$_3$ and b) Sr$_{0.2}$Ca$_{0.8}$RuO$_3$ measured at different magnetic fields.
The curves in b) are shifted by a multiplication of 0.002 m$\Omega\cdot$cm for clarity.
Solid lines were fitted using the $T^2$ law. Arrows mark the temperature $T_{LFL}$ where $T^2$ law starts to deviate from experimental data.}
\end{figure}

\begin{figure}[htb!]
\includegraphics[width=8.5cm]{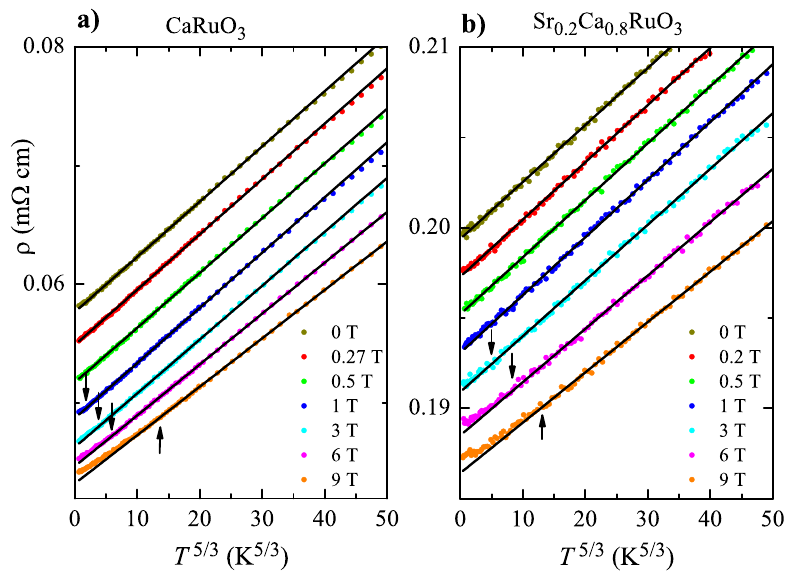}
\caption{\label{f6}
(Color online) 
Resistivity of a) CaRuO$_3$ and b) Sr$_{0.2}$Ca$_{0.8}$RuO$_3$ measured at different magnetic fields.
For clarity, the curves in a) and b) are shifted by a multiplication of 0.003 and 0.002 m$\Omega\cdot$cm, respectively. Solid lines were fitted using the $T^{5/3}$ law. Arrows mark the lower limit of the region where this law well reproduces experimental data.}
\end{figure}

The distinct NFL properties above $T_{LFL}$ and the gradual increase of the range of LFL behaviour 
in higher fields are corroborated by the results of the resistivity measurements. 
The temperature evolution of the electrical resistivity for CaRuO$_3$ and 
Sr$_{0.2}$Ca$_{0.8}$RuO$_3$ are shown in Fig.~\ref{f5} and \ref{f6}. It can be notice that at low temperatures 
the resistivity for each compound demonstrate the LFL temperature dependence (Fig.~\ref{f5}a and \ref{f5}b). 
The least-squares analysis using the formula $\rho(T) = \rho_0 + a_2T^2$ allow to determine 
the residual resistivity $\rho_0$, the field dependent $a_2(B)$ coefficients and to derive the 
$T_{LFL}$ temperatures 
from the upper limit of $T^2$ behaviour.  As it was already mentioned, because of the limited 
accuracy of our measurements such analysis can only be performed when $T_{LFL}$ values exceed 
approximately 2 K.  Fortunately, in the strong magnetic fields range of the Fermi liquid 
behaviour distinctly grows up and even exceeds 8 K for the $x = 0.8$ compound in the field of 9 T.
The values of $T_{LFL}$ derived from the resistivity measurements establish the high temperature part 
of the LFL/NFL phase 
diagram shown in Fig.~\ref{f7}.  In Fig.~\ref{f8} it was proved that the variation of $a_2(B)$ coefficients  with the 
magnetic fields obtained from the data for $B \geq 3$~T fulfil the relation 
$a_2(B)\sim (B - B_0)^{-1}$.\cite{Gegenwart03}  All numerical values of $\rho_0$, $a_2$ and $T_{LFL}$ are 
collected in Table II in the Supplemental material.

\begin{figure}[htb!]
\includegraphics[width=8cm]{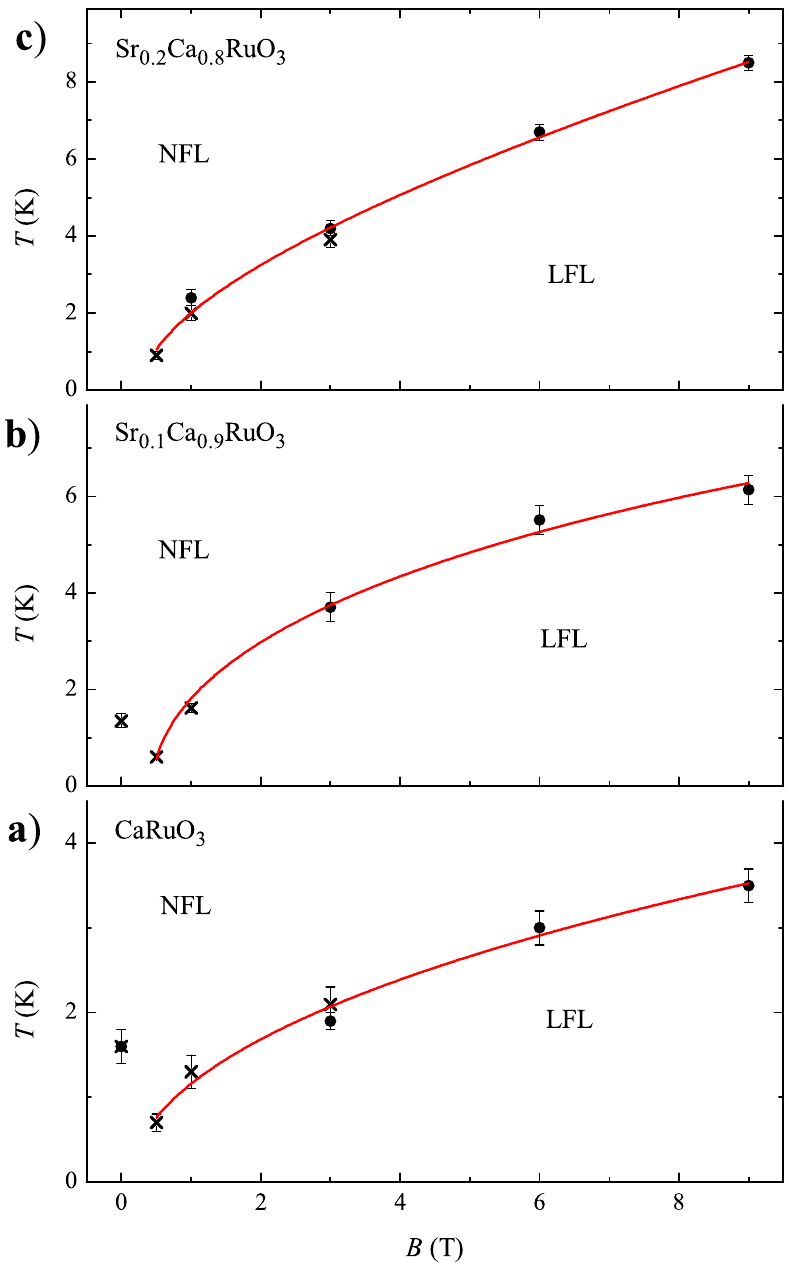}
\caption{\label{f7} $T$-$B$ phase diagrams of Sr$_{1-x}$Ca$_{x}$RuO$_3$ for $x=0.8$, 0.9, and 1.0
with the regions of Landau Fermi liquid (LFL) and non-Fermi liquid (NFL) regions. Points were obtained from analyses of resistivity data (dots) and specific heat (crosses). Solid lines are drawn to guide the eye.}
\end{figure}

\begin{figure}[htb!]
\includegraphics[width=7.5cm]{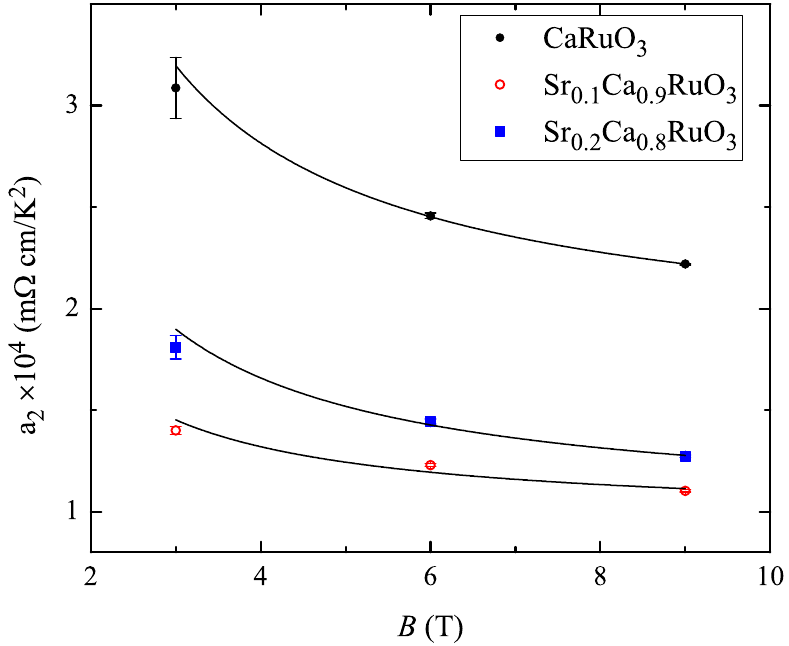}
\caption{\label{f8} Coefficient $a_2$ of the $T^2$ law of the resistivity, determined for Sr$_{1-x}$Ca$_{x}$RuO$_3$ for $x=0.8$, 0.9, and 1.0 in the Landau Fermi liquid phase.}
\end{figure}

As in the zero magnetic field, above the Landau Fermi liquid temperatures $T_{LFL}$ all investigated 
materials show the NFL behaviour characterized by the increase of $C/T$ with decreasing temperature  and  
$T^{5/3}$ dependence of  the electrical resistivity.  The temperature variations of $\rho(T)$ 
for CaRuO$_3$ and Sr$_{0.2}$Ca$_{0.8}$RuO$_3$ in different magnetic fields are shown in 
Fig.~\ref{f6}a and \ref{f6}b.  The continuous lines represent the results of the least-squares analysis using the 
formula $\rho(T) = \rho_0^* + a_{5/3}T^{5/3}$.  One can note that  
in the fields $B_0$ (or close to $B_0$) the NFL relation $\rho(T)\sim T^{5/3}$ very well 
describe the experimental data in all investigated compounds to the lowest accessible temperatures.
It should be also remarked that at or close to $B_0$ this 
NFL formula very well describe the experimental data over one decade of temperature.
The numerical values of the fitted  
parameters ($\rho_0^*$ and $a_{5/3}$ as well as the ranges of application of this description which 
are shown in some drawings by arrows, are given for CaRuO$_3$ and 
Sr$_{0.2}$Ca$_{0.8}$RuO$_3$ in Table II in the Supplemental materials.

\section{Comparison with theoretical models. Discussion.}
In part A of this  section, the experimental results of the specific heat and electrical 
resistivity are compared with predictions of the 
 Moriya's self-consistent renormalization theory (SCR) of spin fluctuations.\cite{Moriya85,Ishigaki96}  
The $C/T$ temperature dependences with subtracted lattice contributions $(C/T)_{lat}$ 
were least-squares fitted using the appropriate SCR formulas for the 3-dimensional 
ferromagnetic materials \cite{Ishigaki96}  and the phenomenological parameters which
characterize the spectrum of spin fluctuations were determined.  Then, these parameters were used 
to analyse the temperature dependence of the resistivity $\rho(T)$.  More details about the whole 
procedure are given 
in the appropriate subsection.  The procedure of the calculation of the lattice 
contribution $(C/T)_{lat}$ is also discussed. In part B, the temperature dependences of 
the electrical resistivity for all investigated compounds and in all applied magnetic fields 
were analysed within the `hidden Fermi liquid' theory of Anderson.\cite{Anderson08} 

\subsection{Specific heat and resistivity within the SCR theory of spin fluctuations}
At the beginning, the self-consistent renormalization (SCR) theory of spin fluctuations developed 
mostly by Moriya  has been successfully applied to describe 
the magnetic and  thermodynamics properties of the 
nearly and weakly ferro- and antiferromagnetic metallic systems.\cite{Moriya73,Moriya85}  Then, this 
theoretical approach was extended to analyse the non-Fermi liquid behaviour of the specific heat 
and electrical resistivity of the antiferromagnetic heavy electron systems in the region of enhanced 
spin-fluctuations near the magnetic instability.\cite{Moriya95}  In particular, the temperature dependence 
the specific heat and electrical resistivity were calculated and compare with the experimental 
data.\cite{Moriya95,Kambe97}  Simultaneously, the theory was applied in the analysis of the anomalous 
non-Fermi liquid behaviour of the specific heat close to the 
magnetic instability in the itinerant ferromagnets and to detailed discussion of the crossover between 
LFL and NFL properties.\cite{Ishigaki96}

In the SCR theory of spin fluctuations the imaginary part of the low frequency 
$\omega$ and long wave-length $q$-dependent dynamic magnetic 
susceptibility $\chi(\omega,q)$ is parametrized by two energy scales $T_0$ and $T_A$ which
characterize the energy width of the spin fluctuation spectrum and the dispersion of the wave-vector 
dependent static susceptibility in the $q$-space, respectively.     
In addition, two dimensionless parameters were introduced: $y_0$ which is a zero temperature inverse 
magnetic susceptibility and measures the proximity to the magnetic instability and $y_1$ 
which reflects the strength of the coupling between the fluctuations with different wave-vector 
(mode-mode coupling).  These four phenomenological parameters which can be experimentally determined 
from the bulk magnetic and neutron scattering experiments allow within the SCR theory to analyse 
the magnetic, thermodynamic and transport properties.  All physical quantities can be derived 
from the reduced inverse magnetic susceptibility $y = 1/(T_A\chi_Q(0))$ ($Q$ is the antiferromagnetic 
order wave-vector, for the ferromagnet $Q = 0$) which can be determined by solution of the 
following integral self-consistent equation
\begin{equation}
y = y_0 + \frac{3}{2}y_1\int_{0}^{1}dx\; x^{3}\left[\ln u - \frac{1}{2u} - \psi(u)\right]  
\end{equation}
where
\begin{equation}
u = \frac{x(y+x^{2})}{t}, \qquad  t = T/T_0, 
\tag{1a}
\end{equation}
and $\psi(u)$ denotes the digamma function.

The temperature dependence of the molar specific heat caused by excitations of the spin 
fluctuations is given by the formula\cite{Ishigaki96}
\begin{equation}
\begin{split}
C_m  = 9N_0k_B\int_0^1 dx\; x^2 \Big\{ \left[u^2-2ux\frac{dy}{dx}+x^2\left(\frac{dy}{dx}\right)^2\right] \\
{}\cdot
\left[-\frac{1}{u}-\frac{1}{2u^2}-\psi'(u)\right]-tx\frac{d^2y}{dt^2}
\left[\ln u-\frac{1}{2u}-\psi(u)\right] \Big\} 
\end{split}
\end{equation}
where $N_0$ is the Avogadro number, $k_B$ is the Boltzmann constant, $\psi'(u)$ denotes the trigamma function and the other parameters 
have the same meaning as in Eq.~(1).
The low temperature limit of Eq.~(2) has the form
\begin{equation}
\begin{split}
\frac{C_m}{T} = \frac{3N_0k_B}{4T_0}\Big[\ln \left(1+\frac{1}{y_0}\right) 
- \frac{2t^2}{5y_0^3}\ln \left(\frac{y_0}{t}\right) \\
+O(t^2)\Big],\quad \text{for } y_0 > 0,
\end{split}
\tag{2a}
\end{equation}
with the saturation value at zero temperature $(3N_0k_B/4T_0) \ln (1+1/y_0)$.  
The temperature dependence at the critical phase boundary is described by the formula
\begin{equation}
\frac{C_m}{T} = \frac{N_0k_B}{2T_0}\ln(1/t),\quad \text{for } y_0 = 0.
\tag{2b}
\end{equation}
This logarithmic behaviour should be observed at low temperatures close to the magnetic instability.

The temperature dependence of the electrical resistivity for the 3-dimensional metal caused by scattering 
of conduction electrons by the ferromagnetic SF is given, within the SCR theory, by the formula\cite{Moriya85}
\begin{equation}
R(T) = r\bar{R}(T),
\end{equation}
where the temperature independent coefficient $r$ allows to distinguish between FM and AFM metals.  The whole 
temperature dependence is contained in the $\bar{R}(T)$ factor which in the case of 3-dimensional ferromagnetic 
metal has the form
\begin{equation}
\bar{R}(T) = 3\int_0^1dx\; x^4\left[-1-\frac{1}{2u}+u\psi'(u)\right]. 
\end{equation}

As in the case of the specific heat calculations the reduced magnetic 
susceptibility $y$ has to be numerically calculated from Eq.~(1). 
At the very low temperatures 
the resistivity is described by the LFL formula $\rho\sim T^2$ and at the critical boundary  
($y_0 = 0$) and in the NFL region one gets\cite{Moriya85}
\begin{equation}
\bar{R}(T) = 0.9385\;t^{5/3}.
\tag{4a}
\end{equation}

Our experimental specific heat and electrical resistivity 
of Sr$_{1-x}$Ca$_x$RuO$_3$ qualitatively follow the theoretical predictions.  At low temperatures 
the specific heat show in general the $C/T = const$ behaviour characteristic for the Fermi 
liquid and at higher temperatures a monotonous pseudo-logarithmic decrease of $C/T$ with 
increasing temperature is observed.  The qualitative agreement with the theoretical predictions 
appears also for the experimental results of the electrical resistivity.  At low temperatures 
characteristic for the LFL $\rho\sim T^2$ behaviour is observed whereas above $T_{LFL}$ 
the NFL variation $\rho(T)\sim T^{5/3}$ is observed.   

In the quantitative analysis one should take into account that the total experimentally determined 
specific heat is the sum of several contributions
\begin{equation}
\frac{C}{T} = \gamma_0 + \frac{C_m}{T} + \frac{C_{lat}}{T},
\tag{5}
\end{equation}
where except the spin fluctuations part $C_m$, $C_{lat}$ denotes the lattice contribution and $\gamma_0$ 
represents the electronic specific heat coefficient which results from the finite density of states 
at the Fermi energy.

The numerical analysis of the specific heat within the SCR theory should in principle give a
proper description of the $C/T$ behaviour and allow to determine the phenomenological 
SCR parameters $y_0$, $y_1$ and $T_0$ was rather complicated and will be described in the following 
in a few steps.  At the beginning of the discussion we concentrate on pure 
CaRuO$_3$ in zero magnetic field where the situation is the most clear.  Then, the analysis is 
extended to specific heat behaviour in external magnetic field and for solid solutions with different 
strontium concentration.  

\begin{figure}[htb!]
\includegraphics[width=8cm]{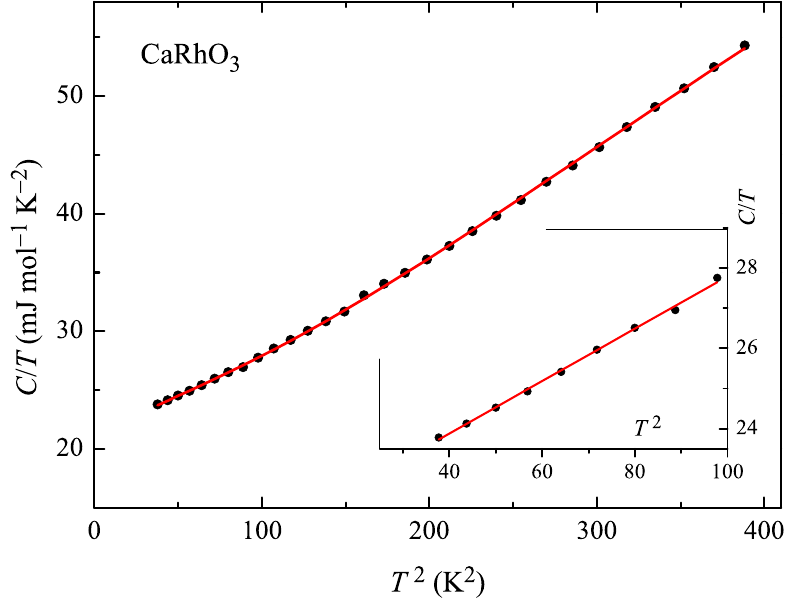}
\caption{\label{f9} Specific heat of CaRhO$_3$ (based on Ref.~\onlinecite{Yamaura09}) 
analyzed within the Debye and Einstein models of the lattice specific heat.}
\end{figure}

Since the NFL anomaly in the temperature dependence of the specific heat is very small 
(increase of $C/T$ ratio below 10 K is below 10 mJ mol$^{-1}$K$^{-2}$) it is very important 
to get independently the reliable temperature dependence of the lattice contribution.  The lattice  
contribution for CaRuO$_3$ used in calculations in this work was obtained from the experimentally 
determined specific heat of CaRhO$_3$ scaled by the procedure proposed in Ref. \onlinecite{Bouvier91}.
CaRhO$_3$ is a metallic paramagnet at least down to 2 K.  The temperature dependence of 
$C/T$ below 20 K does not show any anomalies and can be described as a sum of the 
electronic and lattice contributions.\cite{Yamaura09} The experimental data presented as $C/T$ 
\textit{vs} $T^2$ are shown in Fig.~\ref{f9}. This behaviour was parametrized using the 
Debye and Einstein models of the lattice specific heat.  Such parametrization of $C_{lat}$ should 
be very useful in the analysis of the lattice contributions in the solid solutions where the calcium 
atoms are replaced by the heavier strontium and no suitable analogs with rhodium compounds exist.

\begin{figure}[htb!]
\includegraphics[width=7.5cm]{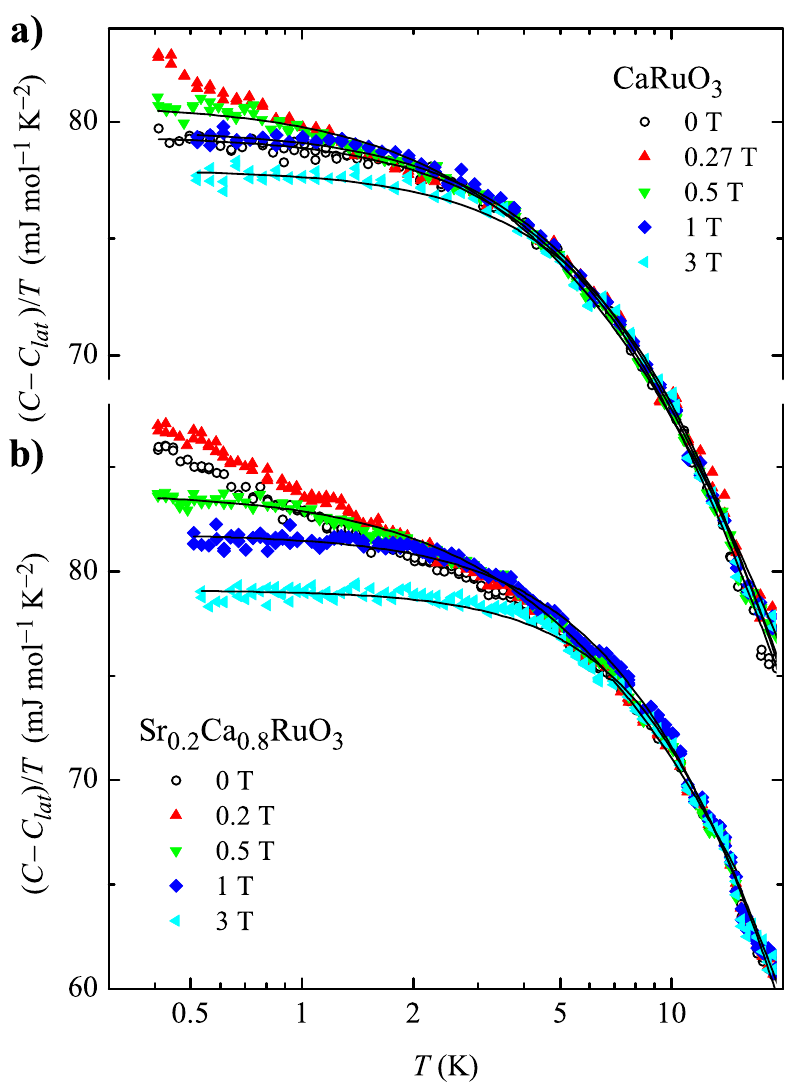}
\caption{\label{f10} 
(Color online) 
Specific heat after subtracting the lattice contribution $C-C_{lat}$ for CaRuO$_3$ and 
Sr$_{0.2}$Ca$_{0.8}$RuO$_3$ at different magnetic fields.
The solid lines were fitted using the SCR model of spin flactuations (see text).
}
\end{figure}

The detailed numerical analysis of the CaRhO$_3$ specific heat proves the linear variation of $C/T$ \textit{vs} 
$T^2$ below approximately 10 K and in this range of temperatures the Debye model is enough to describe the 
specific heat behaviour and the linear fit allows to determine the Debye temperature  $\Theta_D$ 
(Fig.~\ref{f9}, inset).  To get a satisfactory description of $C_{lat}$ above 10 K one has to add 
a small additional contribution which reflects an excitation of optical phonons and is taken into 
account within the Einstein model of the specific heat.  It was found that one frequency parametrized 
as an Einstein temperature $\Theta_E$ is enough to get the very good  description of $C_{lat}$ 
in the range of temperatures up to 20 K. 
The fitted function which contains both contributions is shown as a continuous line in 
the main panel of Fig.~\ref{f9}.
The following values of the parameters were determined from the least-squares analysis: 
$\Theta_D = 530(3)$~K, $\Theta_E = 117(3)$~K and $\gamma_0(\text{CaRhO}_3) = 21.3(1)$ mJ$\cdot$mol$^{-1}\cdot$K$^{-2}$.  The amount of the 
Einstein contribution to the total specific heat is approximately 1.3\%. After the atoms mass 
scaling to CaRuO$_3$ the characteristic temperatures amount $\Theta_D = 533.0$ K and 
$\Theta_E = 117.8$ K.  As a final important remark concerning the 
lattice specific heat one should notice that $C_{lat}$ is completely negligible below approximately 2.2 K, 
so below this temperature the experimental specific heat contains only the electronic and the spin-fluctuations 
contributions. 

After subtraction of the lattice contribution, the specific heat depends on 
four parameters: the electronic contribution $\gamma_0$ and the three parameters 
which describe the spin-fluctuation contribution within the SCR Moriya's theory: $y_0$, $y_1$ and $T_0$. 
In principle, these parameters could be determined simultaneously using the least-squares fitting procedure.  
Nevertheless, a correlation between the parameters 
in connection with the rather smooth experimental $C/T$ curves make the least-squares calculations very 
unstable, for example it is difficult to obtain expected smooth variation of fitted parameters with increasing magnetic field.  More reliable analysis of the $C/T$ behaviour needs reduction of the number of independently 
varied parameters.  The procedure which we used is described below. As in the 
case of determination of $C_{lat}$, the discussion concentrates on pure CaRuO$_3$ in zero field and the rules are later 
applied in the analysis of the specific heat for the other compounds in different magnetic fields. 

First of all, the electronic contribution $\gamma_0$ was determined independently and constrained in the 
further minimization procedure. It was calculated by the analysis of the lowest part of the $C/T$ dependence 
taking into account a correction to the description within the Landau Fermi liquid theory caused by 
the spin-fluctuations.  
The suitable formula which contains the $-T^3\log T$ contribution to the specific heat,    
was obtained within the paramagnon theory by many authors (they are mentioned in the quoted 
reference) and the most convenient form for our purpose is given as\cite{Ikeda91}
\begin{equation}
\frac{C}{T} = 
\gamma_0\left[\frac{m^*}{m}+\alpha_0\left(\frac{T}{T_{sf}}\right)^2\ln\left(\frac{T}{T_{sf}}\right)\right]
\tag{6}
\end{equation}
where $m^*/m$ denotes the mass enhancement factor, $T_{sf}$ is the spin-fluctuation temperature 
and $\alpha_0$ can be expressed by the Stoner enhancement factor.

This formula well describes the $C/T$ ratio even in the wide range of temperatures but 
the values of $\alpha_0$ and T$_{sf}$ strongly 
depend on the temperature range taken into account in the  minimization procedure.  
Nevertheless, $\gamma_0$ and $m^*/m$ which 
are varied as independent parameters do not vary much and for CaRuO$_3$ in zero 
magnetic field $\gamma_0\simeq 10.0(2)$ mJ$\cdot$mol$^{-1}\cdot$K$^{-2}$ and $m^*/m \simeq 7.9(2)$.  
Since the temperature dependent factor 
in formula (6) disappears at zero temperature, the product $\gamma_0(m^*/m)$ 
is equal to the $C/T$ ratio extrapolated to zero Kelvins which for CaRuO$_3$ at zero field 
amounts to $\simeq 79.3$ mJ/molK$^2$.  The values of $\gamma_0$ and $m^*/m$ can be compared with the 
theoretical predictions.  Theoretically calculated value of the electronic contribution 
was reported to be equal to $\gamma_0\simeq 9.54$ mJ/molK$^2$.\cite{Mukuda99}  In addition, 
the detailed theoretical calculations of the electronic structure of SrRuO$_3$ and 
CaRuO$_3$ were recently reported.\cite{Dang15}  Since the values of the mass enhancement factors were calculated for 
different values of an effective onsite interactions $U$ and a Hund's couplings $J$, 
it is not possible to have a direct comparison with the result of our 
analysis.  Nevertheless, our $m^*/m\simeq 7.9(2)$ is in the range of the calculated 
values in some regions of $U$ and $J$ parameters.

Finally, in the first step 
of the analysis  the relation between the spin-fluctuations contribution to the 
specific heat  and the Moriya's parameters at zero kelvins:  
$(C_m/T)_{T=0} = (3N_0k_B/4T_0)\ln(1+1/y_0)$ was used. 
Since $(C_m/T + \gamma_0)$ can be directly determined from the experimental data as the average 
values of $C/T$ in the flat LFL region and with known electronic contribution  $\gamma_0$
one can express $T_0$ by $y_0$ and use only two 
parameters as independent variables in the least-squares analysis: $y_0$ and $y_1$.  
In the last step, when the values of $y_0$, $y_1$ and $T_0$ were in the proper region of the 
parameters space, all 3 SCR parameters were allowed to vary giving the final solution.

Since the described analysis of the electronic and lattice contributions concerned only CaRuO$_3$ 
at zero magnetic field one has to transfer some rules of analysis to compounds with different 
compositions and in the different external magnetic fields.

In order to analyse the specific heat behaviour in different magnetic fields it was 
assumed that the electronic and the
lattice contributions do not vary with field.  The field independence of 
$C_{lat}$ can be inferred from the observation that the experimental $C/T$ values above 15 K 
where the lattice contribution is dominant over the electronic and spin-fluctuations contributions 
practically do not depend on the magnetic field.  Then, the zero temperature 
values of $C/T$ which are given by the product $\gamma_0(m^*/m)$ (see Eq.~(6)) do not 
depend strongly on the magnetic field which means that one should not expect any serious variation of the 
electronic contribution with field.  Indeed, the least-squares analysis of 
the low temperature $C/T$ behaviours using Eq.~(6) shows for each composition practically 
field independent values of $\gamma_0$.  
All of that means that at low temperatures mostly the spin-fluctuations contribution is responsible for the 
observed variation of the specific heat with field. 

As concerns the analysis of the specific heat for compounds with different calcium 
concentration, the zero field $C/T$ behaviour shown in Fig.~\ref{f1} announces serious differences 
in the lattice contributions.  Since as it was already mentioned it would be not 
convenient to determine simultaneously 
all parameters which determine the electronic, lattice and spin-fluctuations contributions the 
analysis was performed in a few steps.  At first, the last-squares analysis of the low temperature 
behaviour using Eq.~(6) shows for each composition practically the same values of 
$\gamma_0 \simeq 10.0(2)$ mJ/molK$^2$.  Then, observation that up to approximately 10 K 
the lattice dynamics is well described by the Debye model, the Debye temperature $\Theta_D$ 
with the SCR parameters and constrained $(C/T)_{T=0}$ was 
calculated by the least-squares analysis of the $C/T$ at low temperatures.  
In the last step, $y_0$, $y_1$ ($T_0$ calculated 
from constrained $(C/T)_{T=0}$) and two additional $C_{lat}$ parameters (the Einstein temperature 
$\Theta_E$ and the Einstein contribution to the total specific heat) were calculated 
with constrained $\gamma_0$ and $\Theta_D$ from the all experimental points up to 20 K.  
The calculations were were performed for the $C/T$ temperature dependences 
for $x = 0.9$ and 0.8 materials in the field 1 T, which show already well developed LFL region.
Finally, it was assumed that as in the case of CaRuO$_3$ for the  
mixed compounds $C_{lat}$ and $\gamma_0$ do not depend on the magnetic field.

As a rule, only the $C/T$ functions with subtracted lattice contributions were 
least-squares fitted within the SCR model using Eq.~(2).  The dimensionless inversed 
magnetic susceptibility $y(T)$ was obtained by numerical solution of Eq.~(1) in a 
self-consistent way.  
The results of these calculations for CaRuO$_3$ and Sr$_{0.2}$Ca$_{0.8}$RuO$_3$ are shown by 
the continuous lines in Fig.~\ref{f10}.  In general, quite good 
agreement between the experimental data and the theoretical curves was obtained for $C/T$ 
behaviour at $B = 0$ (except of $x = 0.2$ sample) and fields $B \geq 0.5$~T.  The results for 
Sr$_{0.1}$Ca$_{0.9}$RuO$_3$ in majority place between results for $x = 0.8$ and 1.0 compounds.
The field dependences of the obtained SCR parameters for different compounds and different 
magnetic fields are shown in Fig.~\ref{f13}.  The numerical values of $y_0$, $y_1$ and $T_0$ were 
gathered in Table I in the Supplemental materials.

Temperature dependences of the specific heat for all materials in weak external magnetic 
fields (0.2--0.3 T) where the $C/T$ ratio increases to the lowest accessible temperatures 
need some additional discussion.  It was found that the increase in $C/T$ values as a function of 
temperature and the maxima in their field dependence can be well descried as a Schottky anomaly 
caused by the ferromagnetic clusters which in the external (or internal for $x = 0.2$ sample in 
$B = 0$) field behave as a two level system.  The possibility of the existence of small magnetic 
clusters which can form at low temperatures the cluster glass state can be inferred from the results 
of the bulk magnetic measurements (difference between FC and ZFC susceptibilities in the weak 
magnetic fields and the hysteresis loop at 2 K) as well as from information delivered by the magnetic 
transition measurements \cite{Demko12} and the inelastic neutron scattering. \cite{Dang15}  
This cluster glass picture allowed for the quantitative description of the $C/T$ behaviour 
as a function of temperature and magnetic field below approximately 2 K.

\begin{figure}[htb!]
\includegraphics[width=8.5cm]{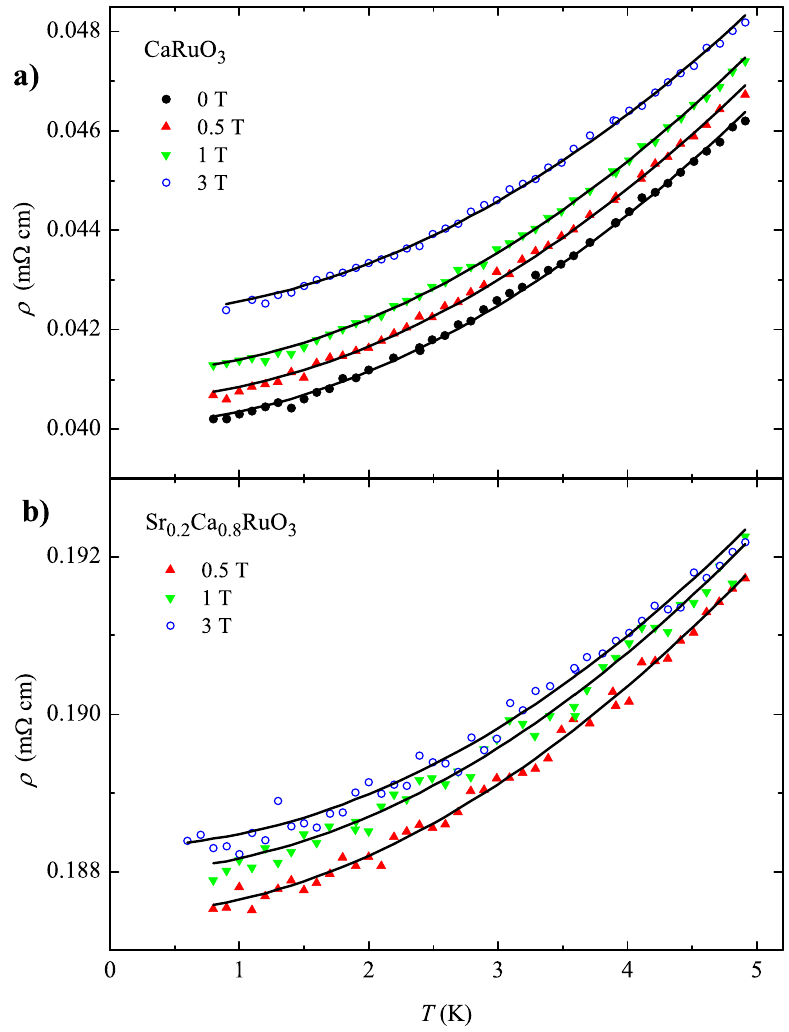}
\caption{\label{f11}
(Color online)  
Resistivity of a) CaRuO$_3$ and b) Sr$_{0.2}$Ca$_{0.8}$RuO$_3$ measured at different magnetic fields.
The curves in a) and b) are shifted by a multiplication of 0.0005 m$\Omega\cdot$cm for clarity.
The solid lines were fitted using the SCR model of spin flactuations.}
\end{figure}

\begin{figure}[htb!]
\includegraphics[width=8cm]{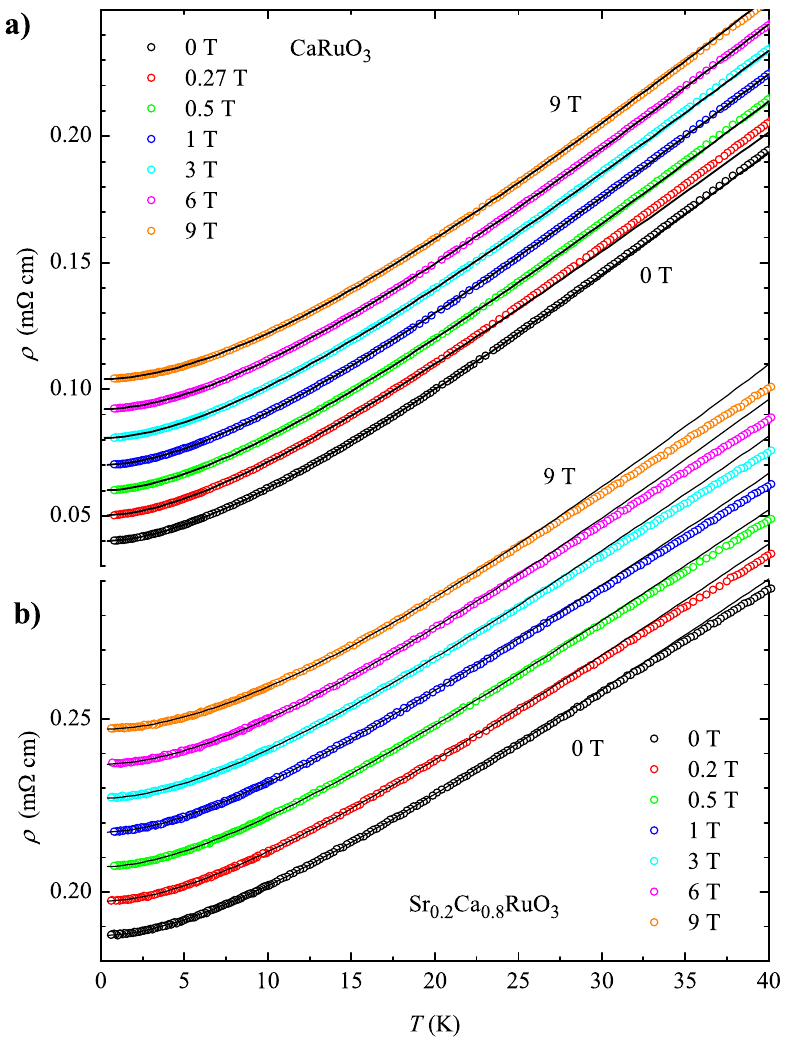}
\caption{\label{f12}
(Color online) 
Resistivity of a) CaRuO$_3$ and b) Sr$_{0.2}$Ca$_{0.8}$RuO$_3$ measured at different magnetic fields.
The curves in a) and b) are shifted by a multiplication of 0.01 m$\Omega\cdot$cm for clarity.
Solid black lines were fitted using the 'hidden Fermi liquid' model of Anderson.}
\end{figure}

In addition, the temperature dependences of the electrical resistivity for all investigated 
compounds in different magnetic fields were analysed within the SCR spin fluctuations theory 
using the formulas (3) and (4).  As in the case of the specific heat analysis the dimensionless 
inverse magnetic susceptibility $y(T)$ was calculated from Eq.~(1). Only the $r$ parameter and the residual 
resistivity $\rho_0$ (not included in Eq.~(3)) were chosen as the variable parameters 
in the least-squares procedure whereas the values of the SCR parameters $y_0$, $y_1$ and $T_0$ 
were constrained to those obtained from the specific heat analysis.  The results of these 
calculations are shown by the continuous lines in Fig.~\ref{f11}.  The analysis shows that quite good 
agreement between the experimental data and theoretical description can be obtained up to 
approximately 5 K.

\begin{figure}[htb!]
\includegraphics[width=8cm]{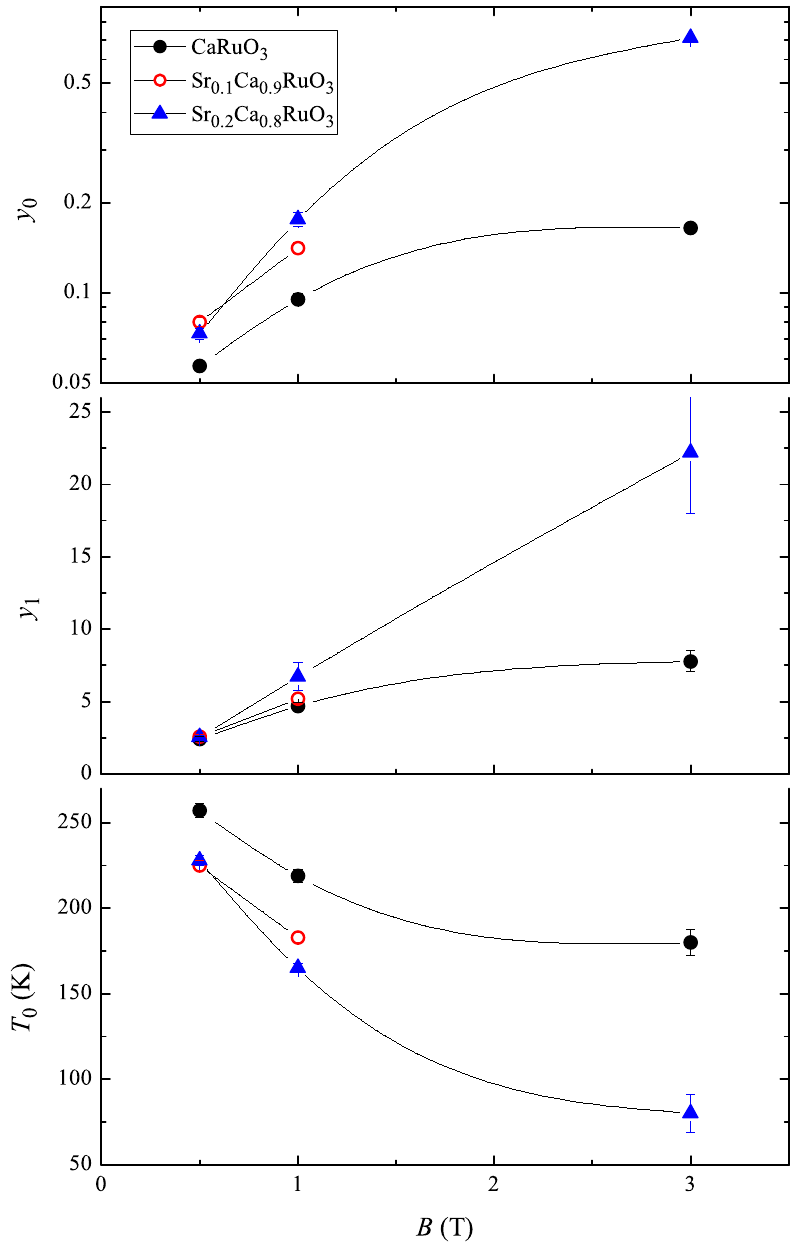}
\caption{\label{f13} 
Magnetic field dependence of the SCR parameters determined from fitting specific heat data
for Sr$_{1-x}$Ca$_{x}$RuO$_3$ for $x=0.8$, 0.9, and 1.0. The solid lines are to guide the eye.}
\end{figure}

\subsection{Resistivity and the `hidden Fermi liquid' theory of Anderson}
A different approach to the problem of NFL behaviour of the electrical resistivity in 
strongly correlated systems was developed 
by P. W. Anderson.\cite{Anderson08}  The aim of his `hidden Fermi liquid' (HFL) theory was the 
explanation of the anomalous electron transport in the normal state of the high-$T_C$ cuprate 
superconductors where the electrical resistivity varies from the linear $T$ dependence in the optimally 
doped region to the LFL behaviour in the overdoped region.  In the HFL theory the 
temperature dependence of the electrical resistivity is described by the following formula 
\begin{equation}
\rho(T) = \rho_0 + a_A\frac{T^2}{T+W_{HFL}},
\tag{7}
\end{equation}
where a prefactor $a_A = \hbar/(e^2E_F)$ and $W_{HFL}$ represents the HFL bandwidth connected 
with the quasi-particles scattering rate which can be probed by the angle-resolved photoemission 
spectroscopy (ARPES).

All the experimental $\rho$ data were analysed using formula (7).  In the least-squares procedure 
the residual resistivity $\rho_0$, $a_A$ prefactor and the $W_{HFL}$ quasi-particles band-width were 
varied as independent parameters.  The results of this analysis for CaRuO$_3$ and 
Sr$_{0.2}$Ca$_{0.8}$RuO$_3$ are shown in Fig.~\ref{f12}.  It can be concluded, that Eq.~(7) assures 
quite good description of the $\rho(T)$ behaviour for the investigated compounds in the wide range of 
magnetic fields up to approximately 25 K.  There is only a very weak field dependence of the $a_A$ 
prefactor which seems to be justified by the formula shown  below Eq.~(7).  On the contrary, the HFL 
bandwidth considerably increases with rising magnetic field (it increases even by a factor of 
two for Sr$_{0.2}$Ca$_{0.8}$RuO$_3$ in the field of 9~T), it means for the materials with increasing 
LFL character.  It seems to be in consent with distinct increase of the $W_{HFL}$ values in the 
La$_{2-x}$Sr$_x$CuO$_4$ high-$T_C$ superconductor with increase of the strontium concentration when the system 
develops from the NFL to the LFL behaviour in the overdoped region.

\section {Short summary}
In this paper we report the results of the specific heat and the resistivity measurements  for 
Sr$_{1-x}$Ca$_x$RuO$_3$ compounds with the calcium concentration $x \geq 0.8$.  The measurements 
were performed in the wide range of temperatures and external magnetic fields.  
In the indicated range of concentrations the materials show the anomalous properties which do not 
agree with predictions of the Landau Fermi liquid theory.  Careful analysis of the $C/T$ and $\rho(T)$ 
behaviour allowed to determine the Landau Fermi liquid temperature which separates the Fermi liquid 
region ($C/T = const$ and $\rho\sim T^2$) from the region of the anomalous non Fermi liquid 
behaviour identified by the  $\rho \sim T^{5/3}$) temperature dependence of the resistivity and 
prepare the $T$--$x$ and $T$--$B$ phase diagrams for the investigated materials.  In addition, the 
experimental results were compared with predictions of the self-consistent renormalization theory 
of spin fluctuations of Moriya (specific heat and resistivity) and with the `hidden Fermi liquid' 
theory of Anderson (resistivity).  Rather good agreement was found between the experimentally 
determined and theoretically calculated behaviour in some ranges of temperature.  Detected anomalous 
increase of $C/T$ below approximately 2 K for all investigated materials at very low magnetic fields 
of 0.2--0.3~T (at zero field for the $x = 0.2$ sample) was interpreted as caused by the Schottky type 
anomaly induced by the ferromagnetic clusters.

\begin{acknowledgments}
This work was financed by the Polish National Science Centre within research project DEC-2011/01/B/ST3/00436.
Measurements were carried out with the equipment purchased thanks to the financial support of the European
 Regional Development Fund in the framework of the Polish Innovation Economy Operational
 Program (contract no. POIG.02.01.00-12-023/08). We thank Prof. K. Yamaura for providing 
us the numerical data of CaRhO$_3$ specific heat.
\end{acknowledgments}

\clearpage

\section{Supplemental material}

\subsection{Bulk magnetic properties of Sr$_{1-x}$Ca$_x$RuO$_3$ ($x \geq 0.6$)} 

The main purpose of the bulk magnetic investigations was to get information about the 
magnetic properties of Sr$_{1-x}$Ca$_x$RuO$_3$ compounds in the range of calcium 
concentration where $T=0$ phase transition between  an itinerant ferromagnet 
and a paramagnetic metal was expected. This critical concentration was obtained by extrapolation 
of the values of Curie temperatures $T_C(x)$ to 0 K as equal to 
$x_c\simeq 0.7$.\cite{Kanbayasi78,Yoshimura98} 
Later on, the existence of inhomogenous ferromagnetic states in some concentration 
range near $x_c$ was discovered by $\mu$SR\cite{Uemura07,Gat11} 
and precise magnetisation measurements on compositionally inhomogenous thin film.\cite{Demko12}

Nevertheless, checking bulk magnetic properties for our samples is important to ensure
that the samples used for thermodynamic and electronic transport investigations 
are of good quality. Then it is important to know exactly 
whether the materials behave as ferromagnets, independently if all Ru moments participate in the 
ordering or there are only ferromagnetic clusters embodied into the paramagnetic matrix, and 
in which materials there is the paramagnetic state without any ferromagnetic impurities.  

\begin{figure}[htb!]
\includegraphics[width=8.5cm]{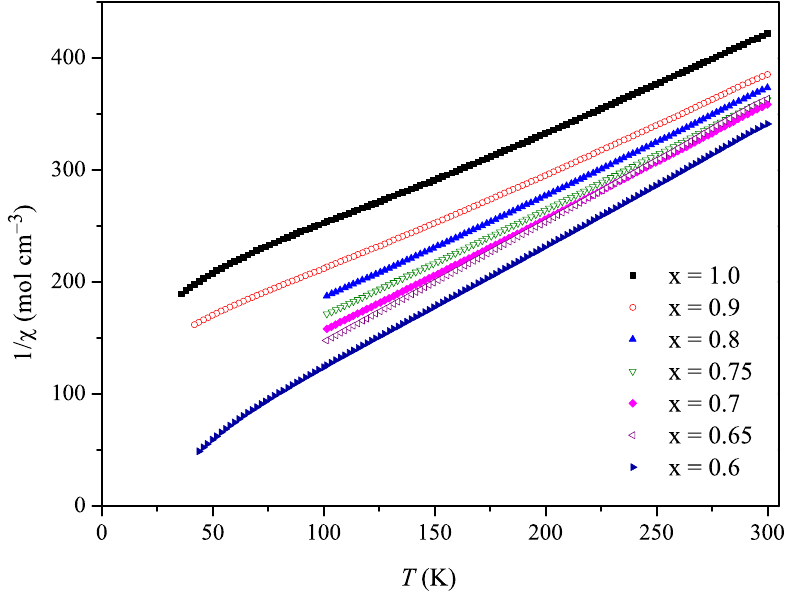}
\caption{\label{fs1} Inverse of magnetic susceptibility measured at 1 kOe for the family of Sr$_{1-x}$Ca$_x$RuO$_3$ compounds.}
\end{figure}

\begin{figure}[htb!]
\includegraphics[width=8.5cm]{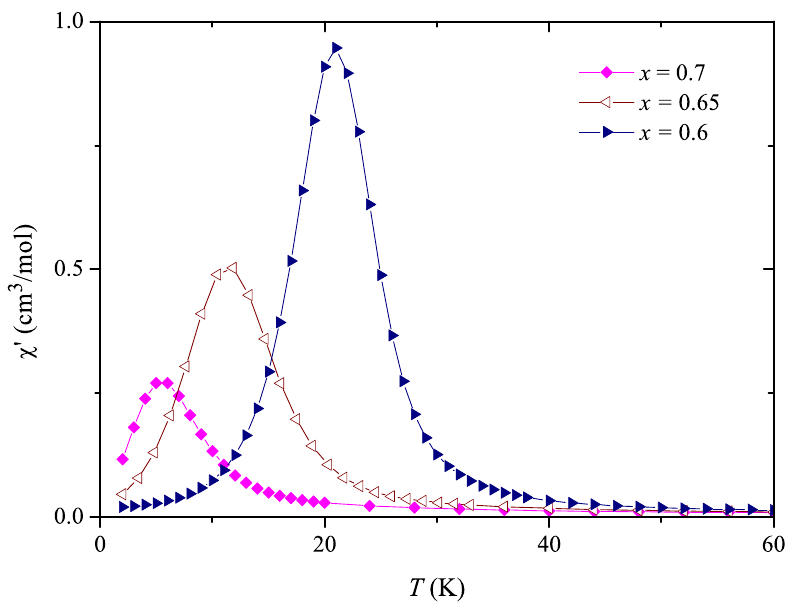}
\caption{\label{fs2} Ac susceptibility measured at 3 Oe, 10 Hz showing ferromagnetic transitions for several Sr$_{1-x}$Ca$_x$RuO$_3$ samples.}
\end{figure}

The bulk magnetic measurements were performed 
on samples  with calcium concentrations $x = 0.6$, 0.65, 0.7, 0.75, 0.8, 0.9 and 1.0 
which cover the regions being on both sides of the critical concentration $x_c$.
The investigations 
included the measurements of the ZFC and FC magnetic susceptibilities in weak magnetic field of 5 mT 
and a series of experiments sensitive to the ferromagnetic ordering which comprised: measurements
of ac susceptibility $\chi'(T)$, search for the hysteresis loop at $T = 2$~K
and the study of the magnetic equation of state (Arrott plot).

Approximately above 100 K, the temperature dependence of the magnetic 
susceptibility for all the investigated materials are well described by the modified Curie-Weiss 
formula $\chi = \chi_0 + C/(T-\Theta_p)$, where $C$ is a Curie constant, $\Theta_p$ 
denotes a paramagnetic Curie temperature and $\chi_0$ contains all the temperature independent 
contributions to the magnetic susceptibility (Fig. \ref{fs1}).  The paramagnetic Curie 
temperature $\Theta_p$ changes sign from positive to negative approximately at 
$x \simeq 0.6$.  It can be also observed that values of the effective magnetic moments 
$\mu_{eff}$ inferred from the values of the molar Curie constants $C = N_0\mu_{eff}^2/(3k_B)$,  
where $N_0$ is the Avogadro number and $k_B$ is the Boltzmann constant, 
practically do no vary with concentration being very close to the value for the ferromagnetic 
SrRuO$_3$ $\mu_{eff} \simeq 2.8\mu_B$, which has probably much more localized character of the 
magnetic moments in the Ru site,\cite{Yoshimura98,Kiyama99,Swiatkowska18}
and show only minor increase for the materials with the calcium concentrations above 0.8.

\begin{figure}[htb!]
\includegraphics[width=8.5cm]{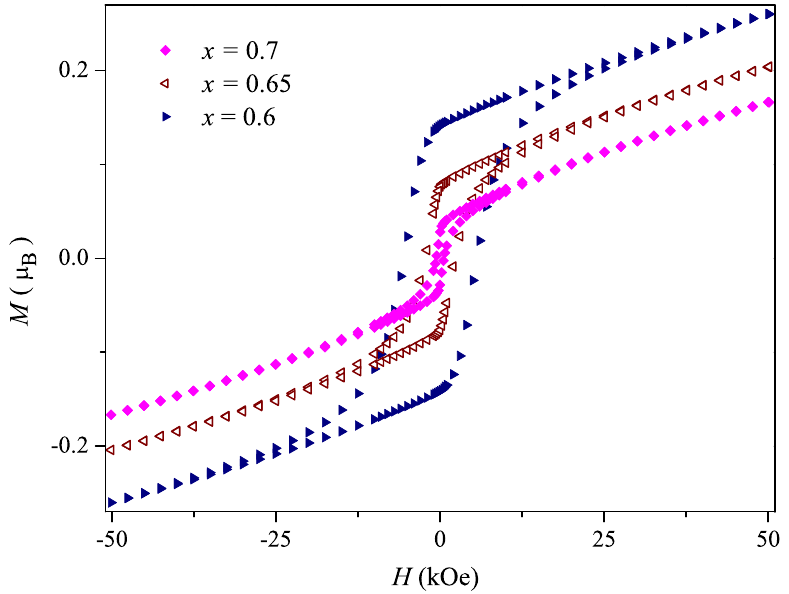}
\caption{\label{fs3} Magnetic hysteresis loops for Sr$_{1-x}$Ca$_x$RuO$_3$ showing ferromagnetic phase.}
\end{figure}

\begin{figure}[htb!]
\includegraphics[width=8.5cm]{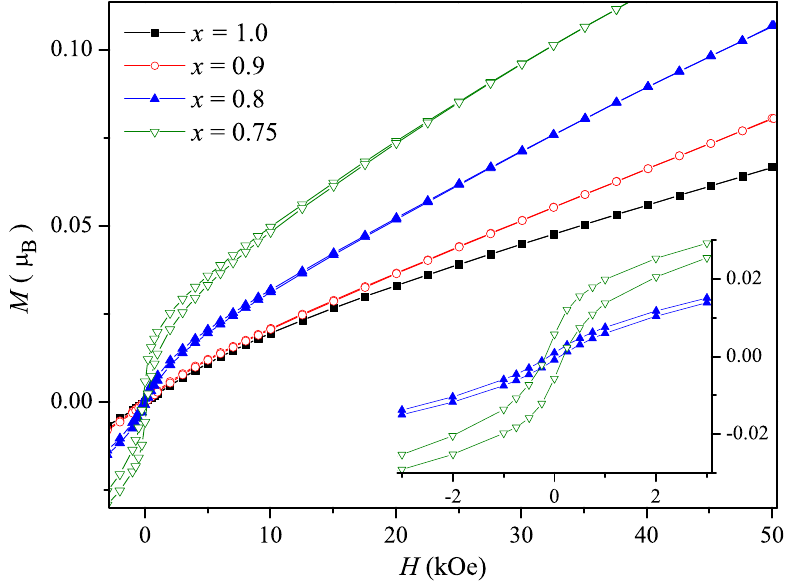}
\caption{\label{fs4} Magnetic hysteresis loops for Sr$_{1-x}$Ca$_x$RuO$_3$ for Ca rich compositions.
Inset: zoomed low field range.}
\end{figure}

The Sr$_{1-x}$Ca$_x$RuO$_3$ compounds with the calcium concentrations $x = 0.6$, 0.65 and 0.70 
behave as ferromagnets.  This is demonstrated by the distinct maxima in the temperature 
dependence of the ac susceptibility (Fig. \ref{fs2}) and by the presence 
of the hysteresis loop at 2 K (Fig. \ref{fs3}). For materials with the calcium concentration 
above 0.70 the hysteresis loop changes the character being much more elongated and very narrow (Fig. \ref{fs4}).
This is probably connected with a partial ordering when only some fraction of the sample 
or even only small clusters show the ferromagnetic behaviour. The narrow hysteresis loops at 2 K 
for the $x = 0.75$ 
and $x = 0.8$ compositions are shown in the inset of Fig. \ref{fs4}. Searching carefully the magnetization 
process one finds also some irreversibility for $x = 0.90$ material an for pure CaRuO$_3$ (not shown).

\begin{figure}[htb!]
\includegraphics[width=8cm]{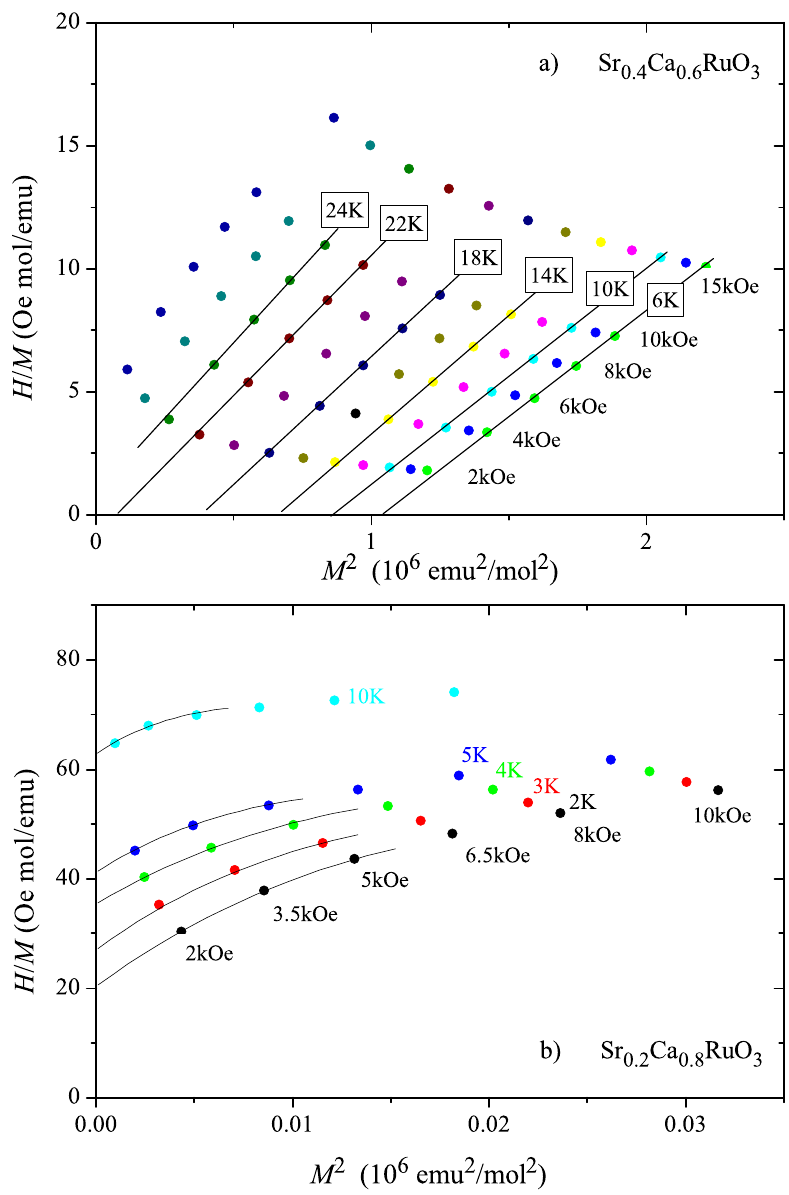}
\caption{\label{fs5} Arrott plot for Sr$_{0.4}$Ca$_{0.6}$RuO$_3$, showing that the sample is ferromagnetic with $T_c$ around 23 K (top) and the similar plot for Sr$_{0.2}$Ca$_{0.8}$RuO$_3$ (bottom) without bulk ferromagnetism present down to 2 K.}
\end{figure}

Different magnetic properties of $x = 0.6$ and $x = 0.8$ are demonstrated 
by the investigation of the magnetic equation of state.  The Arrott plots for these two  materials 
are shown in Fig. \ref{fs5}.  
It is clear that for the sample with $x = 0.6$ there is a different from zero magnetisation at $H = 0$  
at low temperatures. The plot allows to determine from a suitable isotherm 
the Curie temperature $T_{C}\simeq 23(1)$ K.  The Curie temperatures for the samples with 
different calcium concentrations determined from 
the adequate Arrott plots are shown on the $B = 0$ phase diagram in Fig. 3 in the main text.  
Moreover, there is no any isotherm which leads to the different from zero magnetisation at $H = 0$ for the 
Sr$_{1-x}$Ca$_x$RuO$_3$ compounds with $x \geq 0.8$, including pure CaRuO$_3$ (not shown) which means that all 
these compounds are essentially in the paramagnetic state. 
The Arrott plot for the $x = 0.8$ material is shown in Fig. \ref{fs5}(bottom). 
The essentially paramagnetic ground state of CaRuO$_3$, at least 
down to 1.8 K, was also inferred from the results of $^{99}$Ru M\"{o}ssbauer effect 
investigations.\cite{Rams09}   
This of course does not excludes existence of some very small amount of magnetic clusters,
detection of which is below the sensitivity of both methods.

\begin{figure}[htb!]
\includegraphics[width=8.5cm]{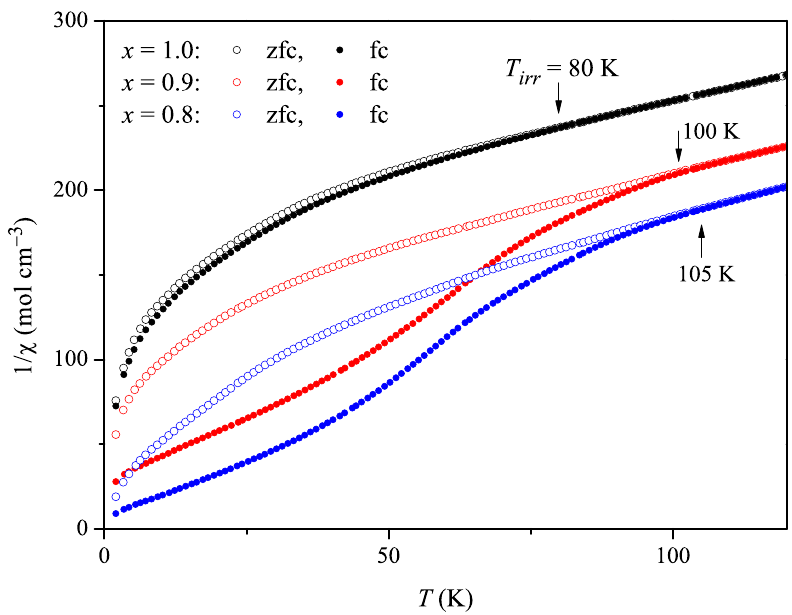}
\caption{\label{fs6} Zero-field cooled (open circles) and field cooled (solid dots) susceptibility measured at
 50 Oe and shown as $1/\chi$, for Ca-rich members of Sr$_{1-x}$Ca$_x$RuO$_3$. }
\end{figure}

As it was already mentioned, to have the proper notion about the magnetic properties of the 
Sr$_{1-x}$Ca$_x$RuO$_3$ compounds,
one has to take into account the results of $\mu$SR investigations.\cite{Uemura07,Gat11}
They show that whereas the compounds with calcium concentration $x \leq 0.6$ are homogeneous 
ferromagnets, which means that all Ru atoms carry magnetic moments which takes part in 
the ferromagnetic ordering.
Materials with $x \geq 0.65$  are magnetically not homogeneous, it means
they contain both magnetic and nonmagnetic fractions.  In addition, the already mentioned 
results of magnetization measurements of thin film\cite{Demko12} show that the ferromagnetic 
phase is extended by the random disorder which could be understand by formation of the 
inhomogenous ferromagnetic material built up of ferromagnetic clusters diluted in the 
paramagnetic matrix.      
Presumably, since the Sr and Ca atoms are randomly distributed in the perovskite lattice, the 
magnetic fraction built up of the ferromagnetic clusters is caused by the regions which 
contain in average more Sr atoms in the nearest surrounding of Ru then anticipated 
from the nominal concentration.   These regions are more favorable for the ferromagnetic order.
Formation of such clusters in some regions of the sample
which are more susceptible to the ferromagnetic order could be the reason of the difference 
between a field cooled (FC) and a zero field cooled (ZFC) susceptibilities in the weak magnetic 
fields of 50 Oe, observed in all investigated samples including pure CaRuO$_3$ (Fig. \ref{fs6}).  
Moreover, the existence of the ferromagnetic clusters in the paramagnetic CaRuO$_3$ was suggested by the results 
of the neutron scattering experiments.\cite{Gunasekera15} In the opinion of the authors of 
this reference such clusters can be induced by the defects in the crystalline structure.

Taking all of that into account one can conclude that even the essentially paramagnetic 
materials which we expect in the high calcium concentration side can have small ferromagnetic 
clusters embodied into the paramagnetic medium.  
Formation of these clusters probably starts in the range of temperature where the zero-field cooled and field cooled susceptibility diverges. The temperature of irreversibility is marked by arrows in Fig. \ref{fs6}.
At high temperatures these clusters are independent and behave as large 
superparamagnetic particles, but at low temperatures even with very weak interaction the system can freeze forming  the cluster glass phase. 

\subsection{$^{99}$Ru M\"{o}ssbauer spectroscopy}

\begin{figure}[htb!]
\includegraphics[width=8cm]{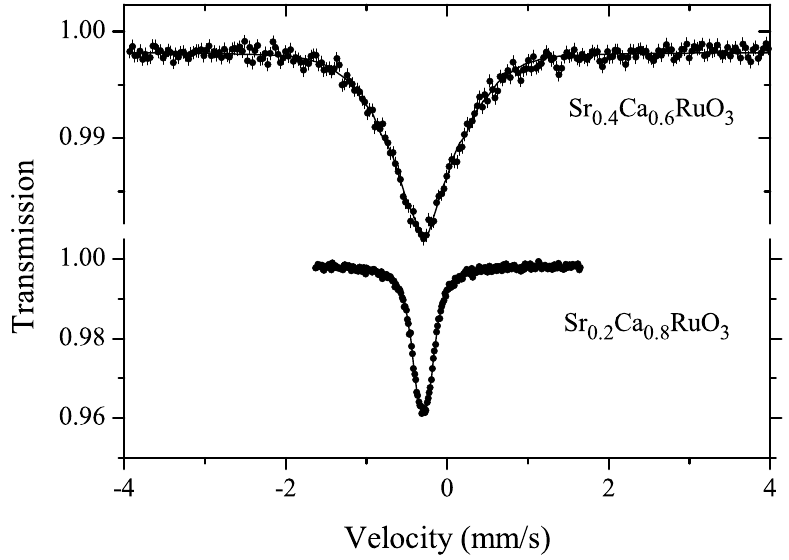}
\caption{\label{fs7} $^{99}$Ru M\"ossbauer spectra for
Sr$_{0.4}$Ca$_{0.6}$RuO$_3$ and Sr$_{0.2}$Ca$_{0.8}$RuO$_3$ at 4.2 K. For the latter no hyperfine magnetic field is measurable, while for the former, significant broadening of the spectrum is caused by magnetic hyperfine field.
}
\end{figure}

The magnetic properties of  Sr$_{1-x}$Ca$_x$RuO$_3$ system were also investigated using 
the $^{99}$Ru M\"{o}ssbauer spectroscopy.  These investigations were performed for 
materials which cover the whole range of calcium concentrations and their results 
will be the subject of a separate publication.\cite{Swiatkowska18}  In this report only 
the M\"{o}ssbauer spectra for materials with the composition $x = 0.6$ and $x=0.8$ are 
presented (Fig. \ref{fs7}).  Both of them were obtained with the $^{99}$Rh source in the matrix 
of metallic ruthenium at $T = 4.2$~K.

Spectrum obtained for the $x = 0.6$ sample was analyzed taking into account the existence 
of the hyperfine magnetic field ($H_{hf}$) at each ruthenium site with  distribution 
of $H_{hf}$ values caused by different surroundings of ruthenium by calcium and strontium 
atoms.  This spectrum confirms that Sr$_{0.4}$Ca$_{0.6}$RuO$_3$ is 
homogeneously magnetically ordered.  In the contrary,  the M\"{o}ssbauer spectra for 
the $x = 0.8$ material was analysed as a nonmagnetic spectrum which, even taking into account 
an extremely small and unresolved quadrupole splitting in the resonance absorber and  
source, looks like a single narrow resonance line.  This spectrum confirms nonmagnetic (paramagnetic) 
state of Sr$_{0.2}$Ca$_{0.8}$RuO$_3$ at 4.2 K.  Nevertheless, this does not contradict 
a possible existence of the ferromagnetic clusters mentioned in the previous section 
with concentration below approximately 3\%.

\subsection{Numerical data inferred from $C/T$ and $\rho(t)$ analysis}
This section contains four Tables of collected almost all results obtained from the analysis 
of the specific heat and electrical resistivity.

\clearpage

\begin{table*}
\caption{\label{t:1}Results of specific heat analysis using the SCR model of T.\ Moriya.}
\begin{ruledtabular}
\begin{tabular}{ccccccc}
compound & $B$ (T) & $T_{LFL}$ (K) & $y_0$ & $y_1$ & $T_0$ (K) & $\gamma_0$ (mJ$\cdot$mol$^{-1}$$\cdot$K$^{-2}$)\footnotemark[1] \\
\hline
CaRuO$_3$ 
& 0 & 1.6(2) & 0.1001(37) & 5.25(26) & 215.4(3.0) & 10 \\
& 0.27\footnotemark[2] & $-$ & $-$ &$-$ &$-$ &$-$  \\
& 0.5 & 0.7(1) & 0.0572(19) & 2.40(17) & 257.0(2.7) & 10 \\
& 1 & 1.3(2) & 0.0954(36) & 4.68(26) & 218.6(3.0) & 10 \\
& 3 & 2.1(2) & 0.1652(14) & 7.77(74) & 179.3(6.2) & 10 \\
\hline
Sr$_{0.2}$Ca$_{0.8}$RuO$_3$
& 0\footnotemark[2] & $-$ & $-$ &$-$ &$-$ &$-$  \\
& 0.2\footnotemark[2] & $-$ & $-$ &$-$ &$-$ &$-$   \\
& 0.5 & 0.9(1) & 0.0727(19) & 2.54(13) & 227.8(2.0) & 10 \\
& 1 & 2.0(2) & 0.1761(84) & 6.71(97) & 164.9(2.0) & 10 \\
& 3 & 3.9(2) & 0.7067(150) & 22.2(4.2) & 79.6(11.0) & 10 \\
\end{tabular}
\end{ruledtabular}
\footnotetext[1]{Parameter fixed during the least-squares fitting
procedure.} 
\footnotetext[2]{Data at this field could not be well reproduced using only this model, and for this reason the
parameters are not given to avoid distorted values.}
\end{table*}

\begin{table*}
\caption{\label{t:t2t53}Resistivity analysis using $\rho=\rho_0+a_2 T^2$ and $\rho=\rho_0+a_{5/3} T^{5/3}$ dependencies.}
\begin{ruledtabular}
\begin{tabular}{cc|ccc|ccc}
compound & $B$ & $T_{LFL}$ \footnotemark[2] & $\rho_0$ & $a_2$ & $a_{5/3}$ & $T$ range \footnotemark[3]\\
   & (T) & (K) & (10$^{-5}$ $\Omega\cdot$cm) & (10$^{-7}$ 
$\Omega\cdot$cm$\cdot$K$^{-2}$) & (10$^{-7}$ $\Omega\cdot$cm$\cdot$K$^{-5/3}$) & (K)\\
\hline
CaRuO$_3$ 
& 0 & 1.6(2) & 3.992(2) & 3.467(6) & 4.646(9) & 1.52--8.75 \\
& 0.27 & $-$ & $-$ & $-$ & $-$ & $-$ \\
& 0.5 & $-$ & $-$ & $-$ & 4.608(7) & below 0.8--9.54 \\
& 1 & $-$ & $-$ & $-$ & 4.646(7) & 1.43--9.51 \\
& 3 & 1.9(1) & 4.067(4) & 3.087(15) & 4.531(7) & 2.22--9.93 \\
& 6 & 3.0(2) & 4.224(4) & 2.457(10) & 4.281(5) & 2.9--10.9 \\
& 9 & 3.5(2) & 4.409(4) & 2.220(4) & 4.075(9) & 4.77--11.6 \\
\hline
Sr$_{0.2}$Ca$_{0.8}$RuO$_3$ 
& 0 & $-$ & $-$ & $-$  & 3.154(17) & below 0.8--9.86 \\
& 0.2 & $-$ & $-$ & $-$  & 3.168(17) & below 0.8--9.70 \\
& 0.5 & $-$ & $-$ & $-$ & 3.154(40) & below 0.8--10.5 \\
& 1 & 2.4(2) & 18.729(8) & 2.58(41) & 3.194(17) & 2.52--10.52 \\
& 3 & 4.2(2) & 18.724(4) & 1.804(58) & 3.084(20) & 3.78--10.55 \\
& 6 & 6.7(2) & 18.715(3) & 1.443(12) & 2.950(20) & 4.53--11.4 \\
& 9 & 8.5(2) & 18.729(3) & 1.274(6) & 2.787(20) & 4.47--11.7 \\
\end{tabular}
\end{ruledtabular}
\footnotetext[1]{Given uncertainties of $\rho_0$, $a_2$, $a_{5/3}$ are statistical errors of fitting using the least-square method. The systematic errors related with the geometry of samples reach 10\%.}
\footnotetext[2]{The maximal $T$ where $T^{2}$ law well reproduces experimental data.} 
\footnotetext[3]{The range of $T$ where the $T^{5/3}$ law well reproduces experimental data.} 
\end{table*}

\begin{table*}
\caption{\label{t:SCR}Results of resistivity analysis using the SCR model $\rho=\rho_0+\rho_{SCR}$ and temperature range from 0.6 to 5 K.}
\begin{ruledtabular}
\begin{tabular}{cccc}
compound & $B$ (T) & $\rho_0$ (10$^{-5}$ $\Omega\cdot$cm) & $a_{SCR}$ (10$^{-5}$ $\Omega\cdot$cm) \\
\hline
CaRuO$_3$ 
& 0 & 4.007(2) & 812.6(8.0) \\
& 0.5 & 4.005(2) & 812.6(5.1) \\
& 1 & 4.012(2) & 824.8(4.0) \\
& 3 & 4.080(2) & 804.0(8.1) \\
& 6 & $-$ & $-$ \\
& 9 & $-$ & $-$ \\
\hline
Sr$_{0.2}$Ca$_{0.8}$RuO$_3$ 
& 0 & $-$ & $-$ \\ 
& 0.5 & 18.743(3) & 505.8(5.8) \\
& 1 & 18.747(4) & 497.3(8.3) \\
& 3 & 18.729(4) & 476.9(8.3) \\
& 6 & $-$ & $-$ \\
& 9 & $-$ & $-$ \\
\end{tabular}
\end{ruledtabular}
\end{table*}

\begin{table*}
\caption{\label{t:anderson}Results of resistivity analysis using 
the Anderson model $\rho=\rho_0+a_A T^2/(T+T_0)$
and temperature range from 0.6 to 24 K.}
\begin{ruledtabular}
\begin{tabular}{ccccc}
compound & $B$ (T) & $\rho_0$ (10$^{-5}$ $\Omega\cdot$cm) & $a_A$ (10$^{-5}$ $\Omega\cdot$cm) & $T_0$ (K)\\
\hline
CaRuO$_3$ 
& 0 & 4.002(2) & 0.5355(9) & 15.7(1) \\
& 0.5 & 4.000(2) & 0.5350(10) & 15.7(1) \\
& 1 & 4.013(2) & 0.5436(9) & 16.3(1) \\
& 3 & 4.069(2) & 0.5430(11) & 16.8(1) \\
& 6 & 4.215(2) & 0.5618(11) & 19.1(1)\\
& 9 & 4.397(2) & 0.5932(14) & 22.8(1) \\
\hline
Sr$_{0.2}$Ca$_{0.8}$RuO$_3$ 
& 0 & 18.747(3) & 0.3491(22) & 14.2(3) \\ 
& 0.5 & 18.730(3) & 0.3499(18) & 14.2(2) \\ 
& 1 & 18.726(3) & 0.3523(24) & 14.4(3) \\ 
& 3 & 18.713(3) & 0.3676(28) & 16.2(3) \\ 
& 6 & 18.693(3) & 0.3955(35) & 20.0(4) \\ 
& 9 & 18.706(3) & 0.4298(50) & 25.2(5) \\ 
\end{tabular}
\end{ruledtabular}
\end{table*}

\end{document}